\documentclass[pra, twocolumn, floatfix]{revtex4-2}
\usepackage{amsmath,amsfonts,amsbsy,amssymb}
\usepackage{sansmath}
\usepackage{mathrsfs}
\usepackage{graphicx}
\usepackage{epstopdf}
\usepackage{url}

\usepackage{ulem}
\usepackage{enumerate}
\usepackage{xcolor}

\usepackage{natbib}

\definecolor{dark_blue}{rgb}{0,0,0.75}
\definecolor{edit}{rgb}{0.8, 0., 0.}

\definecolor{link}{rgb}{0.0, 0.0, 0.0 }
\usepackage[bookmarks = true,
			pdfstartview = FitH,
			colorlinks = true,
			urlcolor=link,
			citecolor=link,
			linkcolor=link,
			hyperfootnotes=false]{hyperref}

\begin{document}
\title{A Quantum trajectory picture of single photon absorption and energy transport in photosystem II}
	
	\author{Robert L. Cook, Liwen Ko,  K. Birgitta Whaley}
	\affiliation{Department of Chemistry, University of California, Berkeley, CA 94720, United States of America}
	
	\date{\today}
	
	\begin{abstract}
        In this work we study the first step in photosynthesis for the limiting case of a single photon interacting with photosystem II (PSII). We model our system using quantum trajectory theory, which allows us to consider not only the average evolution, but also the conditional evolution of the system given individual realizations of idealized measurements of photons that have been absorbed and subsequently emitted as fluorescence.  The quantum nature of the single photon input requires a fully quantum model of both the input and output light fields. We show that PSII coupled to the field via three collective ``bright states'', whose orientation and distribution correlate strongly with its natural geometry.  Measurements of the transmitted beam strongly affects the system state, since a (null) detection of the outgoing photon confirms that the system must be in the electronic (excited) ground state.  Using numerical and analytical calculations we show that observing the null result transforms a state with a low excited state population $O( 10^{-5} )$ to a state with nearly all population contained in the excited states.  This is solely a property of the single photon input, as we confirm by comparing this behavior with that for excitation by a coherent state possessing an average of one photon, using a smaller five site ``pentamer'' system.  We also examine the effect of a dissipative phononic environment on the conditional excited state dynamics.  We show that the environment has a strong effect on the observed rates of fluorescence, which could act as a new photon-counting witness of excitonic coherence. The long time evolution of the phononic model predicts an experimentally consistent quantum efficiency of 92\%.
	\end{abstract}
	
	\maketitle

\section{Introduction}

The light harvesting step of photosynthesis, the initial stage in which photons are absorbed and converted to electrons, is characterized by a remarkably high value of quantum efficiency (QE), reaching values as high as 98-99\% in weak light conditions ~\cite{blankenship_molecular_2014,genty_relationship_1989}. 
Such a highly efficient chemical or physical process is unusual for biological systems, where finite temperature effects and the setting within complex structures that are large on the molecular scale tend to lower the efficiency of dynamical processes.  

This has led to many studies of light harvesting, ranging from biological studies in vivo to ultrafast spectroscopic studies in vitro. 
Time-resolved experiments showing evidence for quantum coherence in excitonic energy transport (EET) have further given rise to a large literature of both experimental~\cite{brixner_two-dimensional_2005,engel_evidence_2007} and theoretical~\cite{wang_quantum_2019,cao_quantum_2020} studies focused on the nature of the excitonic states accessed after absorption of light, the relative roles of electronic and vibrational degrees of freedom, and the manifestation of dynamical coherences in ultra-fast experiments employing laser excitation of photosynthetic systems.  These studies are often motivated by the desire to understand whether dynamical coherences in EET are in any way essential or responsible for the high quantum efficiency in vivo.  

Focusing entirely on the nature of excitonic energy transport bypasses a second key aspect of light harvesting, namely the absorption process, which arises when we consider the microscopic meaning of the QE.  The QE is usually stated and interpreted as the efficiency of conversion of a single {\em absorbed} photon to a single electron or electron-hole pair~\cite{blankenship_molecular_2014}. However, determining in a non-destructive way that a single quantum system absorbed a particular incident photon is not a trivial task. The QE is typically inferred by measuring the relative change in {\em fluorescence} under low and high intensity light exposure~\cite{blankenship_molecular_2014}.  These experiments provide only a macroscopic average estimate of the QE, based upon bulk measurements of a macroscopic ensemble and under a rate equation limit.  
\begin{figure}[htbp]
\includegraphics[width=0.9\columnwidth]{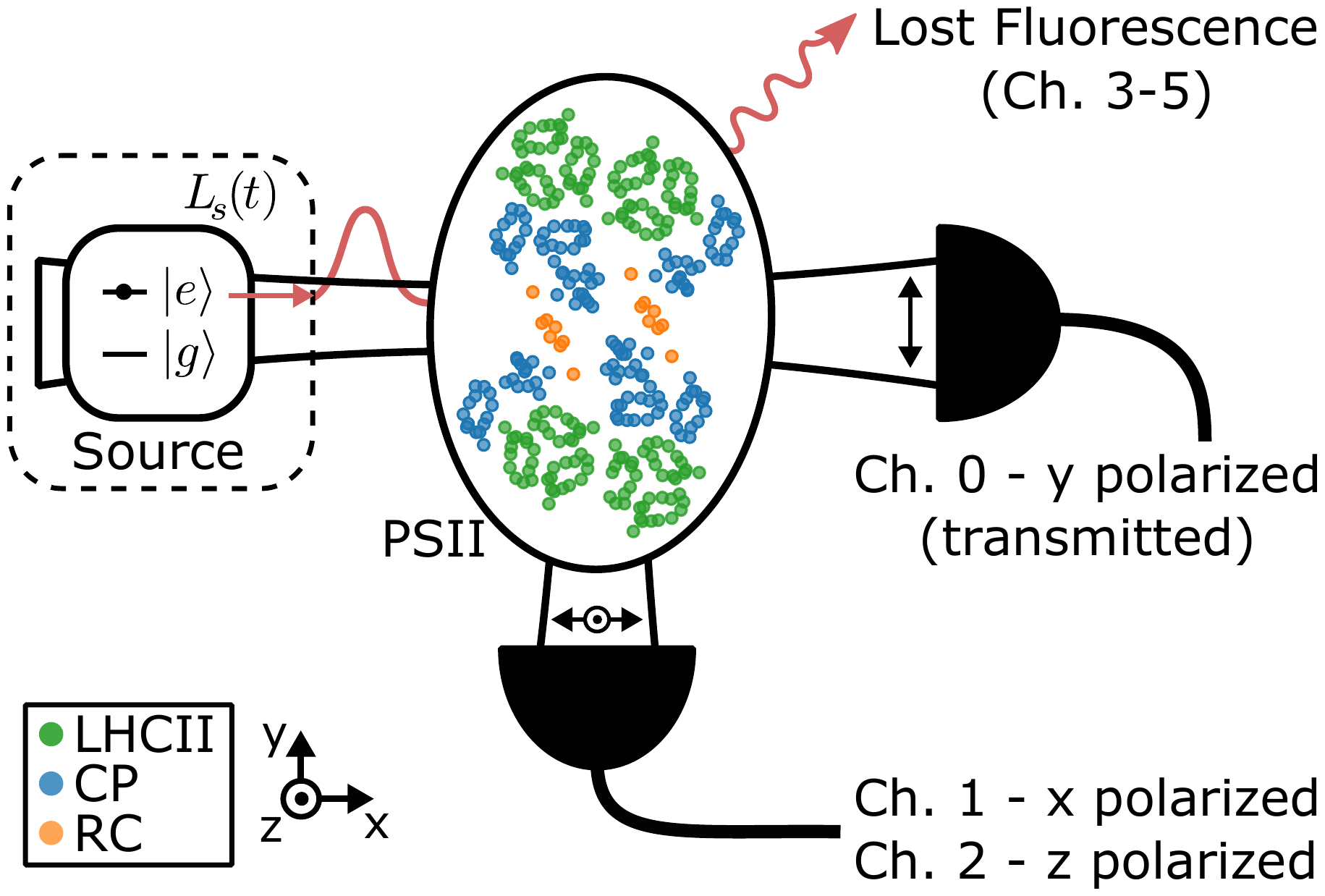}
\caption{\label{fig:schematic} Single photon interaction schematic: A model source produces a single photon in a $\mathbf{e}_y$ polarized traveling wave packet of a known shape, incident upon PSII. The transmitted signal is measured by an ideal photon counter, channel 0.  Fluorescence in an orthogonal spatial mode can be measured with $\mathbf{e}_x$ ($\mathbf{e}_z$) polarization in ch. 1 (ch. 2). A complete accounting of all photon field coupling requires additional loss chs. 3-5 for source polarizations $\mathbf{e}_x$ - $\mathbf{e}_z$, respectively, which propagate into unobserved directions.}
\end{figure}

What is needed in order to gain a more complete understanding of light harvesting and its high QE are  microscopic studies of single photons incident on light harvesting systems that correlate the released single electrons with an {\em independent} measure of photon absorption.  In particular, one needs to be able to verify that absorption of a photon has occurred, a difficult task given the low excitation probability of the photosynthetic pigments~\cite{chan_single-photon_2018}.

The most straightforward method for independently verifying the absorption of a photon is to utilize a single photon counter and look for a reduction in the number of photons transmitted through the sample.  This is illustrated schematically in Fig. \ref{fig:schematic}, with the PSII complex of green plants taken as a prototypical photosynthetic system.  Here one and only one photon is known to be produced by a source which interacts with the light harvesting supercomplex photosystem II (PSII). Such a source can be realized by a spontaneous parametric down-converting crystal (SPDC) producing photon pairs that are used to generate heralded single photons~\cite{kaneda_heralded_2016}.  If a perfectly efficient photon counter fails to detect a transmission signal after a herald photon is detected (channel 0 in Fig. \ref{fig:schematic}), then the photon must have been absorbed.  The fraction of these events that also produce electron-hole pairs, were these to be measured in a time delayed coincidence with the herald photon, as implemented in studies of vision in~\cite{phan_interaction_2014}, will then give a direct measure of the quantum efficiency. 

We note that for general states of light, the failure to detect an outgoing photon in channel 0 can occur in one of two ways. In addition to the possibility of photon absorption mentioned above, the second option is that there was no incoming photon in the first place.  A perfectly heralded source of down converted photons from a SPDC, removes the second option. However, when the heralded single photon source is replaced by a source of weak classical light, i.e., a laser or blackbody radiation with an average photon number $\bar{n} \ll 1$, both options exist with finite probability.  Moreover, the second option of no incoming photon is much more probable than the first option of absorption. We will show that this has a dramatic impact upon the probability that the system is excited, giving a very different outcome than that from a true single photon.

In this work we construct a tractable theoretical model that accurately accounts for the quantum mechanical nature of the photon absorption and subsequent possible emission of a fluorescent photon at the same or a smaller energy. We focus our attention on the initial absorption event populating a bright excitonic state, together with the subsequent EET to a chromophore of equal or lower energy that can fluoresce.  Ideally this fluorescent chromophore is located close to the reaction center sites from which single electrons are typical released and the fluorescence can then be taken as a proxy for electron-hole pair formation.  We note that experimental techniques for carrying out ultrafast spectroscopy on molecular systems with photocurrent detection are being developed~\cite{bakulin_ultrafast_2016} and we envisage that these would eventually be combined with controlled single photon incidence, as has been done in studies of vision~\cite{phan_interaction_2014}, although realizing the combination of these two disparate techniques for photosynthetic systems poses significant challenges in practice.

The current paper primarily focuses on the pigment electronic degrees of freedom, i.e., the excitonic modes, and develops a quantum trajectory theory for the interaction with single photons in the context of PSII.  However, it is known that a crucial component in energy transport is the effect of coupling between the excitonic degree of freedom to its vibrational environment~\cite{ishizaki_quantum_2010}.  In order to study the vibrational contribution, we compare two extreme cases.  The first is the fully coherent model where the excitonic system is completely isolated from its environment.  The second is a fully Markovian model of the exciton-phonon interaction, where the average effects of the phonon bath ultimately results in classical hopping dynamics between energy eigenstates.  In the context of quantum information, the phonon bath results in a unobserved decoherence channel, whose loss of information necessitates a mixed state description.  A subsequent paper will extend the trajectory picture to include measurements of the phononic modes, and thereby recapturing all information that is in principle available to the outside observer~\cite{cook_phonons_2022}. 

In addition to the phononic decoherence, we also  include a coarse description of radical pair formation and non-radiative decay.  This expanded phononic model enables us to make a prediction of the microscopic QE, given our independent measure of single photon absorption. This yields a QE value of 92\%, consistent with experimental estimates from macroscopic kinetic studies~\cite{blankenship_molecular_2014,genty_relationship_1989}. 

Our approach to study the quantum dynamics of a light harvesting system interacting with the input and output photon fields due to single photons utilizes the formalism of quantum trajectory theory,
in which the outgoing light fields are considered to be experiencing continuous quantum limited measurements~\cite{carmichael_quantum_2000}.  In the absence of these measurements, the Markovian interaction between the system and the photon fields would lead to a decoherence process and description of the system dynamics would require a mixed density matrix picture. The power of quantum trajectory theory is that all decoherence channels are assumed to be the result of measurements that are `recorded' by an observer or environment, and therefore the system remains in a single pure state throughout.  However, this time dependence of this pure state now depends upon the random measurement outcomes at each time, and so its evolution is given by a stochastic Schrodinger equation.  While the quantum trajectory theory is often used merely as a convenient Monte Carlo computational tool~\cite{plenio_quantum-jump_1998}, this picture is perfectly suited for investigating questions about single photon absorption, since we can explicitly compare individual trajectories for which photon counts are observed in channel 0 of Fig. \ref{fig:schematic}, to trajectories for which no photon counts are observed.  

In this analysis, each photon count in any channel corresponds to (induces) a discrete quantum jump from the optical active ``fluorescent'' states of the system to its electronic ground state.  If the photon detection occurs in a channel other than 0, that photon must have originated from the system and thus the count indicates that the system has fluoresced and is now in its ground state.  In contrast, a detection in channel 0 can occur via two indistinguishable options or ``paths".  The first is through fluorescence while the second option is that incident photon was never absorbed in the first place and is merely being transmitted.  The quantum electric fields for these two processes add in superposition and so a photon counter cannot unambiguously distinguish between these two possibilities. However, fluorescent photons that are temporally and/or spectrally well separated from the incident pulse will be essentially distinguishable, particularly when filters are employed.

Regardless of the source of the photon, i.e., whether from fluorescnence or from transmission, when a quantum of energy is detected in ch. 0 we know that this energy is no longer in the system. The system must therefore be in the ground state.  Such quantum trajectories provide an ``unraveling" of the average (unconditioned) dynamics of the system under the continuous measurement to give a ``conditioned" dynamics.  For Markovian baths such as the radiation field from which single photon states are sampled, the resulting conditioned dynamics are accepted as having a valid physical interpretation (see, e.g., the discussion in \cite{plenio_quantum-jump_1998}).   

The remainder of the paper is structured as follows.  Sec. \ref{sec:qualitative} describes the qualitative features of our model in more detail and discussed the role played by the measurement channels.  
Sec. \ref{sec:emission} then considers the quantitative details of the system-light interaction. 
In particular we derive here the appropriate jump operators for interaction of a photosynthetic complex with a general light field, given a specific structure for the complex that specifies the relative positions and orientations of all chromophores.  
Sec. \ref{sec:SSE} extends the analysis to include the role of the incident photon, derives the relevant equation, and examines the characteristic superpositions, the so called ``bright states'' that are created by absorption of the incident photon. Sec. \ref{sec:simulations} then outlines our numerical simulations for PSII and describes the quantum trajectories that result from these and their implications for understanding the fluorescent photon distributions obtained from absorption of single photons.  We also use a smaller five site chromophore model (pentamer) to demonstrate the difference  between the dynamics conditioned on detection of single fluorescent photons obtained from a single photon input and the corresponding dynamics obtained from a weak coherent laser. Sec. \ref{sec:phonons} then considers the coupling to a phononic environment, defining a Markovian model of the phonon coupling and including radical pair formation and non-radiative loss.  We then present numerical simulations showing how the decoherence introduced by phonons changes the fluorescent count rates, and compute a final probability for radical pair formation which provides an estimate of the QE.  Finally, in Sec. \ref{sec:conclusion} we conclude with a summary and a discussion.

\section{The Model \label{sec:qualitative}}
Fundamental to the nature of quantum mechanics is the fact that the act of measuring a quantum system almost always results in a dramatic change in its future evolution.  On average (or when the measurement results are not recorded), these changes manifest as a strongly decohering process that competes with the original dynamics. This is usually described by evolution of the density matrix of the system. However, measurements made on a system's Markovian environment will not disturb the system any further and thus analysis of the measurement outcomes via the environmental degrees of freedom can provide partial information about the state of the system.  

In particular, the quantized light field is well modelled as a Markovian bath~\cite{loudon_quantum_2000} where the quantum fluctuations on the incoming field are completely uncorrelated on the shortest time scale relevant to the absorbing system.  Any light propagating away from the system remains causally disconnected from this and so any operations performed on that light cannot increase the decoherence felt by the system.  Thus by continuously measuring the system's radiation field environment, an external observer gains all of the available information afforded by quantum mechanics without further disturbing its average dynamics.  This is not to say that the measurements are independent of the system.  Indeed some measurement outcomes result in individual ``quantum trajectories" of the system - specified by a particular measurement record - being remarkably different from the average evolution~\cite{plenio_quantum-jump_1998}.

\subsection{Correlating fluorescent photons with discrete quantum jumps}
How this information is obtained by the environmental degrees of freedom depends upon how the measurement is performed and thus on the nature of the measurement record.  Here we shall consider idealized photon counting measurements and thus the record will indicate the time and location of the detected photon. The discontinuous nature of this measurement will induce discontinuous changes in our state of knowledge about the system, i.e., causing the classic ``quantum jump'' on photon emission.  However, in the situation that
a photon was known to be incident upon the system and was not detected, then it must have been absorbed.  This knowledge can also induce a kind of quantum jump.  A quantum jump induced by such a null result plays an important result in this work. Such jumps have been discussed previously~\cite{ruskov_crossover_2007} and were also utilized in a recent feedback scheme to control and reverse the jump~\cite{minev_catch_2019}. 

Specifically, consider the idealized case of a single photon incident upon an absorbing system, e.g., PSII, as illustrated schematically in Fig. \ref{fig:schematic}.  A single photon is generated in a time varying wave packet (details of this generation are given in Sec. \ref{sec:SSE}), and is incident upon the target system, PSII. On arrival at the system, the incoming beam has linear polarization along the $\mathbf{e}_{y}$ direction.  The transmitted part of the incident beam is collected and measured by an ideal photon counter in channel 0.  Without the absorbing system, the single photon input will generate one and only one count in this detector.  However, in the presence of the absorbing system, a failure to detect the input photon in channel 0 indicates that the incident single photon must have been absorbed somewhere.  In the ideal case the only absorber present is the target system, so this null result in channel 0 implies that the system must be excited.

The key fact is that without the measurement in channel 0, the system and field states will evolve into an entangled superposition which has a small amplitude ($|\alpha|^2 \ll 1$) of having the system in excited states of PSII with no photons in the field, and a large amplitude for ground state PSII with one transmitted photon, i.e., 
\begin{equation}
    |\psi \rangle = \alpha\,|\text{vac} \rangle |\text{excited}\rangle  + \sqrt{1 -  |\alpha|^2 }\, |\text{ch. 0 photon} \rangle |\text{gnd} \rangle , 
\end{equation}
where we have written the state of the field in a number representation first, and the state of PSII second. If we average over the photon field, then the majority of the resulting mixed state of PSII is in the ground state with probability $1- |\alpha|^2$, and the remaining small fraction is in the excited state.  However, if we measure an outgoing photon in channel 0, then we destroy this superposition and project the state $|\psi \rangle$ onto the ground state.  Expressed in the language of information, if we measure a photon in channel 0, then we know the incident excitation is not in PSII and thus must be in its ground state.  Conversely, by not measuring a photon in channel 0, our certainty that the system must have absorbed the photon is reflected in the fact that the once small probability to be in any excited state will now be re-normalized to be one, with zero probability to be in the ground state.  

This simple picture becomes more complicated when we consider the continuous time arrival of the incoming field.  At intermediate times there is then a third possibility, namely that the incoming field is still excited and the system has yet to interact with it.  This complication means that a null result observed in channel 0 for only a finite time window conveys only partial information, leaving open the possibility that the system may still be in its ground state.  The continuous gain of this information has the effect of smoothing out the transition.  We shall return to this point in Sec. \ref{sec:Bayes}.

A definitive verification that an excitation did occur can be made by putting a second detector located off-axis from the first one.  Suppose the input field propagates along $x$, with polarization  $\mathbf{e}_y$. We then consider two additional channels, 1 and 2, that detect the two orthogonal polarizations, $\mathbf{e}_x$ and $\mathbf{e}_z$, for a single spatial mode propagating along $y$, see Fig.~\ref{fig:schematic}. If the excited system decays via photon emission, there is then a finite probability for that emitted photon to be counted in channels 1 or 2.   Because there is only one photon, only one of the three detectors in channels 0 - 2 will respond in any given individual experimental realization.  Thus a count in any one detector will be perfectly correlated with null results in the other two detectors. 

Furthermore, in the absence of vibrations photons that are resonantly absorbed and then spontaneously emitted will typically have a delay that is on the order of the radiative lifetime $\approx$ 10 ps.  In addition to the value of this average decay time, the quantitative details of the excited state dynamics also have an effect upon when this decay occurs for an individual photon.  To understand how this arises, in section \ref{sec:fluorescence} we analyze the Fourier spectrum of the time dependent decay and show how this is directly related to these excited state dynamics.  

\subsection{The role of bright states in absorption}
In section \ref{sec:bright_states}, we show that the linearity of the electric dipole interaction as well as the fundamentally quantum nature of the single photon state implies that the excited system state, i.e., the bright state, is in principle a superposition state over multiple chromophores, with the extent of the superposition dependent on the coupling between the transition moments of the chromophores, which depend in turn on their relative locations and orientations.  This bright state superposition is generally distinct from the energy eigenstates, since it derives from the total interaction of all chromophores with light, rather than from the excitonic Hamiltonian.  It is also strongly dependent on the orientation of the chromophore dipole moments relative to the incident photon polarization. Sec.~\ref{sec:phonons} shows how such a coherent excitation is modified when the system is in the presence of a strongly dephasing Markovian phononic bath.

A general criterion for determining when collective excitation occurs is that all chromophores be separated by a distance that is much smaller than an optical wavelength.  This is the same criteria for superradiance, i.e., collective photon emission from an ensemble of radiators~\cite{dicke_coherence_1954}.  In superradiance, the rate at which photons are emitted from the ensemble depends on both the number of radiators involved, and the number of excitations present in the system.  Here we are predominantly interested in the absorption of photons rather than their emission.  

The collective absorption of a single photon is a well studied phenomenon since the seminal work of~\citeauthor{dicke_coherence_1954}~\cite{dicke_coherence_1954}.  Several proposals for long distance quantum communication~\cite{hammerer_quantum_2010, duan_long-distance_2001} use atomic ensembles to store and deterministically retrieve single photon states.  For atomic ensembles the collective nature of the excitation serves as a natural protection against some sources of decoherence~\cite{hammerer_quantum_2010,gorshkov_universal_2007}  and is also accompanied by collective emission, i.e., superradiance. However, excitations in photosynthetic systems are very different from the collective excitations of atomic ensembles because of the presence of a phonon bath. While there is some evidence for a limited amount of initial delocalization following optical excitation in some systems~\cite{novoderezhkin_exciton_1999, savikhin_excitation_1998,malina_superradiance_2021, yakovlev_exciton_2002}, fluorescence studies have shown that the steady states at longer time are localized in only a few chromophores~\cite{monshouwer_superradiance_1997,kuhn_pumpprobe_1997}.

Here we investigate what role the vibrational degrees of freedom play, both during and subsequent to absorption of a photon. We do so in two extreme limits.  The first is the baseline of a phonon free evolution, where the conditional dynamics are that of a coherently evolving bright state.  The second is that of a Markovian model where each chromophore interacts with an independent phononic bath. This simple phononic model is commonly used to study energy transport in these photosynthetic systems~\cite{ishizaki_quantum_2010, kuhn_pumpprobe_1997,mohseni_environment-assisted_2008, plenio_dephasing-assisted_2008}, which result in the commonly applied Bloch-Redfield equations.

We limit our investigation here to a single PSII complex for two reasons.  First and foremost is that this allows for tractable numerical simulations. The second reason is that the coupling between multiple PSII complexes is expected to be negligibly small under in vivo conditions~\cite{amarnath_multiscale_2016}.  Thus even if a single photon could in principle create a superposition across many complexes, that superposition would play no role in the subsequent energy transport. This is in stark contrast to the bright state superposition created within a given complex. Here the local Hamiltonian will take the internal superposition created by the photon and transform it into another delocalized state within the same complex.  

We use the same specific model of PSII as that of Ref. ~\cite{bennett_structure-based_2013}, for which the Hamiltonian was constructed using published values for the on-site energies and site-to-site couplings with in a given sub-complex, together with a dipole approximation for the coupling of chromophores across sub-complexes~\cite{bennett_structure-based_2013}.  We note that this model is an approximation to the true structure and energies, in particular with the reaction center (RC) from a cyanobacterium being used as a proxy for the RC in higher plants.

Our model of PSII contains a total of 324 chlorophyll pigments. These are arranged into 24 sub-complexes containing 8 to 16 chromophores, with the entire supercomplex showing a pronounced bilateral symmetry (Fig.~\ref{fig:schematic}).  Towards the exterior are the trimers formed by the light harvesting complex II (LHCII): these chromophores are shown in green in Fig.~\ref{fig:schematic}. Further in the interior are a total of 10 (5 per dimer) core antenna complexes (CP), shown in blue.  The final charge separation occurs at one of the two central reaction centers (RC), which are shown in orange. 

\section{Deriving the collective emission operators\label{sec:emission}}

Here we consider explicitly only the Q$_\text{y}$-band transitions in the chlorophyll pigments, which in PSII have energies $\sim 15,300 \text{ cm}^{-1}$ ~\cite{bennett_structure-based_2013}. The fundamental electromagnetic coupling between the various chromophores in PSII has two distinct contributions. The first of this is the well-known dipole--dipole (van der Waals) interaction, while the second is the collective emission of freely propagating radiation, which is often referred to as superradiance ~\cite{gross_superradiance_1982}.  While such collective quantum optical effects are generally considered at ultra low temperatures or in the context of ultracold atomic gases ~\cite{dicke_coherence_1954, hammerer_quantum_2010,duan_long-distance_2001}, a limited form of collective emission has been observed in the bacterial light harvesting systems LH1 and LH2~\cite{monshouwer_superradiance_1997} and also recently in the bacteriochlorophyll c aggregates in chlorosomes of green photosynthetic bacteria~\cite{malina_superradiance_2021}.

The system coupling to the freely radiating electromagnetic field is treated under the Born--Markov approximation, which is valid for a weakly coupled system.  For the case of multiple radiating sites, this requires the same assumptions as those leading to the Wigner--Weisskopf spontaneous emission rate for a single atom.  Working in second order perturbation theory and with the radiation field assumed to be in vacuum and at zero temperature results in the master equation~\cite{lehmberg_radiation_1970}, 
\begin{widetext}
\begin{equation}\label{eq:superME}
\frac{d\rho}{dt} =-i [H, \rho] 
+ \sum_{i,j} \gamma_{ij} \left( \hat{\sigma}_{-}^{(j)} \rho\, \hat{\sigma}_{+}^{(i)} - \tfrac{1}{2} \hat{\sigma}_{+}^{(i)} \hat{\sigma}_{-}^{(j)}\, \rho - \tfrac{1}{2} \rho\, \hat{\sigma}_{+}^{(i)}\hat{\sigma}_{-}^{(j)} \right),
\end{equation}
where $H$ is the Hamiltonian that governs the electronic coupling, $\hat{\sigma}_{\pm}^{(j)}$ is the atom raising/lowering operator for site $j$, and $\gamma_{ij}$ are the spontaneous decay rates.  The decay rates are given by~\cite{carmichael_quantum_2000},
\begin{equation}\label{eq:gamma_ij}
\begin{split}
\gamma_{ij} = \gamma_0\, \frac{3}{2} &\left[ \left( \frac{\sin(k_0\, r_{ij}) }{k_0\, r_{ij} } + \frac{\cos(k_0\, r_{ij} ) }{(k_0 \,r_{ij} )^2} -  \frac{\sin(k_0\, r_{ij}) }{ (k_0\, r_{ij})^3 } \right) \boldsymbol{d}_{i} \cdot \boldsymbol{d}_{j} \right. \\
&\quad - \left. \left(\frac{\sin(k_0\, r_{ij}) }{k_0\, r_{ij} } + 3 \frac{\cos(k_0\, r_{ij} ) }{(k_0 \,r_{ij} )^2} - 3 \frac{\sin(k_0\, r_{ij}) }{ (k_0\, r_{ij})^3 } \right)  \frac{\boldsymbol{d}_{i}\cdot \mathbf{r}_{ij}\,  \mathbf{r}_{ij} \cdot \boldsymbol{d}_j }{{r_{ij}}^2} \right],
\end{split}
\end{equation} 
\end{widetext}
where $\mathbf{r}_{ij} = \mathbf{r}_i - \mathbf{r}_j$ the vector distance between pigments $i$ and $j$,  $r_{ij} = |\mathbf{r}_i - \mathbf{r}_j|$, $k_0$ is the resonant wave number, $\mathbf{d}_i$ is the (unitless) transition dipole moment, and $\gamma_0$ the (vacuum) Wigner--Weisskopf decay rate for a dipole of unit strength, i.e., for $d_0 = 1 \text{ Debye}$.  

For PSII, we have typical interchromophore distances $r_{ij} \lesssim 30 \text{ nm} $, and a characteristic wave number representing an average over chromophore excitation energies of $k_0 \sim 2\pi/ (650 \text{ nm} )$.  Thus we are safely in the regime where $k_0\, r_{ij} \ll 1$, which allows for the point dipole approximation to Eq.~(\ref{eq:gamma_ij}):
\begin{equation}\label{eq:gamma_ij_approx}
\gamma_{ij} \approx \gamma_0\, \mathbf{d}_i \cdot \mathbf{d}_j.
\end{equation}

In order to account for the fact that the photosynthetic system is embedded in a non-absorbing dielectric medium, we will extend the above master equation assuming an ``empty cavity model''~\cite{glauber_quantum_1991}.  In such a model, the medium with index of refraction $n_r$ forms a weak optical cavity around the system, which results in an enhancement of the vacuum decay rate by a local field correction factor.  In this work, we assume an index of refraction close to that water ($n_r = 1.33$) and will use an in-medium decay rate per unit dipole strength given by 
\begin{equation}\label{eq:gamma_0}
    \gamma_0 = n_r \left(\frac{3 n_r^2}{2 n_r^2 + 1}\right)^2\, \frac{d_0^2 \omega_0^3}{3 \pi \epsilon_0 c^3}.
\end{equation} 
For water, this results in an increase of in-medium decay rates by 80\% over the corresponding Wigner-Weisskopf rates in vacuum. We take the transition dipole moments of the chromophores, ${\bf d}_i$ from the model of \citeauthor{bennett_structure-based_2013}, namely within the RC and the interior complexes CP47, CP43, the chlorophyll a chromophores have transition dipole moment magnitudes of 4.4 Debye, while the remaining chlorophyll a in the light harvesting complex II (LHCII) and minor complexes have transitions moment magnitude 4.0 Debye.  All chlorophyll b chromophores have transition moment magnitude 3.4 Debye. In the reaction center, the pheophytin molecules have transition moment magnitude 3.5 Debye. The dipole orientations for the RC and interior complexes were extracted from the protein data structure derived from~\cite{umena_crystal_2011} and the LHCII and remaining minor complexes were extracted from~\cite{liu_crystal_2004}.  See ref~\cite{bennett_structure-based_2013} for further details about the structural model.

A quantum trajectory unraveling of the superradiant master equation was first given in ref. \cite{carmichael_quantum_2000}.  In that work, the outgoing radiation was partitioned into a large number of differential areas, with each differential solid angle representing a distinct jump channel, i.e., a photon emitted in that direction. The average over this differential partitioning was then shown to generate the desired master equation. 

Here we take a more minimal approach, in which we first identify the smallest number of jump operators that are needed in order to recreate the appropriate average.  We show below that since $k_0\, r_{ij} \ll 1$, the smallest number of jump operators is three, corresponding to three distinct collective dipoles radiating in orthogonal directions in space.  From this minimum number, we then identify jump operators that correspond to the radiation collected by our three detectors, which take an identical form to the differential operators derived in Ref. \cite{carmichael_quantum_2000}.  Finally we arrive at a complete set of jump operators after deriving the collective jump operators that correspond to emission of photons by the collective dipoles into directions not captured by our detectors.  These represent the loss channels.  Combining the three detection modes with one loss channel for each of the three collective dipoles, gives a minimum number of six required jump operators.  

\subsection{Mode decomposition\label{sec:modes}}

We first rewrite the master equation, Eq.~(\ref{eq:superME}), in Lindbad form. This is generally just an algebraic manipulation and results in an equivalent master equation.  However it will allow us to identify the natural collective emission dipole operators. Specifically, we work with the collective vector dipole operators
\begin{equation}
    \mathbf{D} \equiv \sum_{i = 1}^N \mathbf{d}_i\, \hat{\sigma}_{-}^{(i)}.
\end{equation}
The utility of these operators becomes evident by first noting that the (scalar) operator $\mathbf{D}^\dag \cdot \mathbf{D}$ is equal to  
\begin{equation}
    \mathbf{D}^\dag \cdot \mathbf{D} = \sum_{i,j = 1}^{N} \hat{\sigma}_{+}^{(i)} \mathbf{d}_i \cdot \mathbf{d}_j  \hat{\sigma}_-^{(j)}. 
\end{equation}
Then under the approximation $\gamma_{ij} \approx \gamma_0 \mathbf{d}_i\cdot \mathbf{d}_j$, the final term in Eq.~(\ref{eq:superME}) can be compactly rewritten as
\begin{equation}
    -\sum_{ij} \frac{\gamma_{ij} }{2} \rho\, \hat{\sigma}_{+}^{(i)} \hat{\sigma}_-^{(j)} \approx - \tfrac{\gamma_0}{2} \rho\, \mathbf{D}^\dag \cdot \mathbf{D}.
\end{equation}
Similarly, Eq.~(\ref{eq:superME}) can be shown to be equal to
\begin{equation}
    \frac{d\rho}{dt} =-i [H, \rho] + \gamma_0 \mathbf{D} \cdot (\rho \mathbf{D}^\dag ) - \tfrac{\gamma_0}{2} \mathbf{D}^\dag \cdot \mathbf{D}\, \rho - \tfrac{\gamma_0}{2} \rho\, \mathbf{D}^\dag \cdot \mathbf{D}.
\end{equation}

A Lindblad decomposition of this equation is now easily arrived at by writing the dot products in terms of a set of basis vectors $\{\mathbf{e}_i : i \in 1,2,3 \}$ and defining the operators
$\tilde{L}_i = \sqrt{\gamma_0}\, \mathbf{e}_i\cdot \mathbf{D}$, so that
\begin{equation} \label{eq:Lindblad}
    \frac{d\rho}{dt} =-i [H, \rho] + \sum_{i = 1}^3 \left( \tilde{L}_i \,\rho\, \tilde{L}_i^\dag - \tfrac{1}{2} \tilde{L}_i^\dag \tilde{L}_i\, \rho - \tfrac{1}{2} \rho\,\tilde{L}^\dag_i \tilde{L}_i \right).
\end{equation}
When describing the master equation, the choice of basis is irrelevant, since a different basis would yield a different Lindblad decomposition but leaves the sum in Eq.~(\ref{eq:Lindblad} invariant. However, when considering a specific unraveling of that master equation, different decompositions correspond to physically distinct measurements.  So although how one chooses to measure the environment cannot change the {\it average} evolution of a Markovian system, the information gained from the measurement and therefore how the state is updated conditioned on that information clearly does depend upon the measurement specifics.

For the analysis of light absorption and emission by photosynthetic complexes by photon counting experiments, we introduce the measurement jump operators 
\begin{equation} \label{eq:measure_jump_ops}
    L_p \equiv \sqrt{\gamma_0\, \eta }\, \mathbf{e}_p \cdot \mathbf{D}, 
\end{equation}
where $\eta < 1$ is a unitless coupling efficiency factor discussed below, and $\mathbf{e}_p$ is the polarization vector of the measured electric field component at the sample location.  In Fig. \ref{fig:schematic} we show three ``physical'' measurement channels labeled by $p = 0,1,2$.  Channel 0 describes a photon counter located in the far-field such that it measures photons propagating nominally in $+\mathbf{e}_x$ direction, with linear polarization parallel to the $\mathbf{e}_y$ axis.  Channels 1 (2) measure photons propagating in the $-\mathbf{e}_y$ direction with polarization aligned to the $\mathbf{e}_x$ ($\mathbf{e}_z$) axis, respectively.

In ref.~\cite{carmichael_quantum_2000}, each jump operator had an over all efficiency factor of $\eta = \frac{3}{8 \pi} \Delta \Omega,$ where $\Delta \Omega$ is the small solid angle collected by the detector.  In appendix \ref{app:paraxial} we derive an identical coupling expression using a paraxial mode decomposition.  

In order to accurately reproduce the Lindblad form of Eq.~(\ref{eq:Lindblad}), we must also derive additional Lindblad jump operators that correspond to the emission of photons into unobserved directions.  The jump operators ${L}_p$ are no longer sufficient since they already include the photons emitted into the detected channels. Thus we seek a new set of operators $\{ L_{\perp\, i}  \}$, such that the constraint
\begin{equation}
    \gamma_0\, \mathbf{D}^\dag \cdot \mathbf{D} = \sum_{i} L_{\perp\, i}^\dag L_{\perp\, i} + \sum_{p = 0}^{2} L_p^\dag L_p
\end{equation}
is satisfied.  This constraint has a trivial solution for the case of the geometry in Fig. \ref{fig:schematic}, i.e., when all three Cartesian axes are equally 
represented by the measurement operators $L_p$. In appendix \ref{app:geometry} we derive a general expression for $\{ L_{\perp\, i} \}$, for a measurement geometry that involves non-orthogonal measurement directions.  Substituting Eq.~(\ref{eq:measure_jump_ops}) and rearranging terms, the constraint becomes
\begin{equation}\label{eq:ch_loss_constraint}
\begin{split}
        \sum_{i} L_{\perp\, i}^\dag L_{\perp\, i} &= \gamma_0\, \mathbf{D}^\dag \cdot \mathbf{D} - \gamma_0 \eta \sum_{p = 0}^{2} \mathbf{D}^\dag \cdot \mathbf{e}_p\, \mathbf{e}_p \cdot \mathbf{D} \\
        &= \gamma_0 (1 - \eta) \mathbf{D}^\dag \cdot \mathbf{D}.  
\end{split}
\end{equation}
An obvious solution to the constraint is then given by 
\begin{equation}
    L_{\perp\, p} = \sqrt{\gamma_0\,(1 - \eta)}\, \mathbf{e}_p \cdot \mathbf{D}.
\end{equation}
We note that because our measurement operators preserve the spherical symmetry of the over all dissipation, any orthogonal basis for $L_{\perp\, p}$ would be sufficient.  In appendix \ref{app:geometry} we derive a general expression for $\{ L_{\perp\, p} \}$, for a measurement geometry that involves non-orthogonal measurement directions.

\section{Single Photon Stochastic Schrodinger Equation\label{sec:SSE} }
Here we utilize a stochastic Schrodinger equation to unravel the Lindblad master equation, Eq.~(\ref{eq:Lindblad}), For such Markovian systems, the resulting quantum trajectories are determined by the choice of measurement operators and may be interpreted as representing the evolution of individual physical trajectories~\cite{plenio_quantum-jump_1998, wiseman_quantum_2010}. Such an unraveling of a given master equation evolves under piece-wise deterministic equations of motion interrupted by discontinuous jumps and is known as a jump unraveling.  The equation of motion is for an unnormalized state vector, $\tilde{\psi}$,
\begin{equation}\label{eq:unnormalized_dpsi}
    \tfrac{d}{dt}\tilde{\psi} = (-i H - \tfrac{1}{2}\sum_{j} L^\dag_j L_j ) \tilde{\psi} \equiv -i H_\text{eff}\, \tilde{\psi}.
\end{equation}
Here the index $j$ runs over all jump channels, i.e., both the observed measurement channels $j = 0, 1, 2$, as well as the fluorescence loss channels $j = 3,4,5$.  Within each infinitesimal time increment $\Delta t$, the probability for detecting a jump in channel $j$ is 
\begin{equation}\label{eq:prob_j}
    \mathsf{prob}_j(t) = \langle \tilde{\psi}| L_j^\dag L_j |\tilde{\psi} \rangle \Delta t.
\end{equation}
Upon detecting a jump in channel $j$, the state undergoes the discontinuous transformation:
\begin{equation}
    \tilde{\psi} \mapsto L_j \tilde{\psi} / \| L_j \tilde{\psi} \|.
\end{equation}
The stochastic Schrodinger equation is related to the density matrix of the Lindblad evolution by the convergence 
\begin{equation}
    \rho(t) = \lim_{N \rightarrow \infty} \frac{1}{N} \sum_{m = 1}^N \frac{| \tilde{\psi}_m(t) \rangle \langle \tilde{\psi}_m(t) |}{\|\tilde{\psi}_m(t)\|^2},
\end{equation}
where each $\tilde{\psi}_m(t)$ is a quantum trajectory generated by an independently sampled measurement record. 

The most common single photon sources are derived from heralded nonlinear optical processes, in particular from pairwise two-photon production by spontaneous parametric downconversion (SPDC)~\cite{uren_generation_2005,kaneda_heralded_2016}.  While in general the photon statistics from such a source can be non-trivial due to residual time and frequency entanglement, here we assume that all such entanglement has been removed and the joint spectral amplitude factorizes.  This ensures that the down converted photons can be assumed to be in a tensor product, so that the detection of the heralding photon merely confirms the knowledge that the incident photon is propagating in a well defined spatial mode with a known wave packet envelope function $\xi(t)$, and with no knowledge of the entanglement of the original photon pair produced by SPDC. 

Given this knowledge that a single photon is incident upon the system with a well defined carrier frequency, $\omega_0$,  polarization at the sample $\mathbf{e}_0$, and time dependent envelope function $\xi(t)$, the details of how that photon was created are then no longer relevant for our analysis. In particular, we note that such a photon could just as easily been created by an arbitrary two level system which couples exclusively to the relevant spatial mode, with this coupling modulated to generate the desired envelope. Indeed, Ref.~\cite{gough_quantum_2011-1} has shown that an incoming single photon can be simulated by a cascaded model in which an upstream excited qubit is coupled to the downstream photon system via a time dependent decay operator $L_s(t)$.  In order to reproduce a specific $\xi(t)$, the decay operator must take the form 
\begin{equation}
    L_s(t) = \frac{\xi(t)}{\sqrt{w(t)}} \sigma_-,
\end{equation}
where the weighting factor
\begin{equation}
    w(t) \equiv \int_t^\infty |\xi(s)|^2 ds
\end{equation}
is equal to the amount of the incoming wave packet that has yet to arrive at the system at time $t$. Thus $w(t)$ is the probability that an idealize photon counter will detect the photon in the time interval $[t, \infty)$.  

This single qubit source model provides a convenient minimal description of the single photon. Thus if the qubit is in the excited state $|e\rangle$, then the photon is in the incoming, upstream field and has not yet arrived at the system.  If the qubit is in the ground state, $|g\rangle$, then there is no longer an incoming photon.  The photon has either been absorbed by the system or it has continued on to the field downstream from the system, where it can be lost to the environment or measured at a photon counter.

In absence of a downstream absorbing system of chromophores, the single qubit unnormalized state vector $|\tilde{\psi}_q \rangle$ evolves under the equation
\begin{equation}
    \frac{d}{dt} |\tilde{\psi}_q \rangle = - \frac{1}{2} \frac{|\xi(t)|^2}{w(t)} \sigma_+ \sigma_- |\tilde{\psi}_q \rangle.
\end{equation}
For the initial condition $|\tilde{\psi}_q (t_0)\rangle = |e\rangle$ at $t_0$, with $\int_{t_0}^\infty |\xi(t)|^2 dt = 1$ for a single photon wave packet, this equation has the solution
\begin{equation}
    |\tilde{\psi}_q(t) \rangle = \sqrt{w(t)} |e \rangle,
\end{equation}
which is easily verified by direct substitution. 

In the presence of the downstream chromophore system, this constitutes a cascaded quantum system~\cite{gardiner_quantum_2004, gough_quantum_2011-1}, where the upstream single photon emitter now has the effect of modifying both the Lindblad jump operators as well as the coherent Hamiltonian describing the interaction of the photon field with the chromophore system.  Suppose the output of the qubit emitter is uniquely routed to the input channel $0$, with corresponding downstream jump operator $L_0$.  The jump operator for the combined output from both the photon source and system emission into the ultimate output of channel $0$ is then
\begin{equation}
    L_\text{tot}(t) = L_s(t) + L_0.
\end{equation}
Taking the possible absorption of the upstream photon into account leads to the modified system Hamiltonian,
\begin{equation}\label{eq:H_cascade}
    H_\text{tot}(t) = H_\text{sys} -\tfrac{i}{2} L_s(t) L_0^\dag + \tfrac{i}{2} L_s^\dag(t) L_0.
\end{equation}
Combining the new emission operator $L_\text{tot}(t)$ with $H_\text{tot}(t)$ to construct the cascaded non-Hermitian Hamiltonian then results in the cancellation of the last term in Eq.~(\ref{eq:H_cascade}), yielding
\begin{multline}\label{eq:H_eff_cascade}
    H_\text{eff}(t) = H_\text{sys} - i L_s(t) L_0^\dag
    - \tfrac{i}{2} L_s^\dag(t) L_s(t) - \tfrac{i}{2} \sum_{j = 0} L_j^\dag L_j.
\end{multline}

We note that for a one dimensional channel, the dynamics of the downstream system should have no effect on the upstream qubit.  So for an initial condition where the photon qubit is in $|e\rangle$ and the downstream system is in its ground state $|gnd\rangle$, the joint qubit-system unnormalized state vector should always be in the following form:
\begin{equation}\label{eq:unnorm_psi_tot}
    |\tilde{\psi}_\text{tot}(t) \rangle = \sqrt{w(t)} |e\rangle |gnd \rangle + |g\rangle |\beta(t) \rangle.
\end{equation}
Here $|\beta(t)\rangle$ is an unnormalized state vector in the excited system subspace, i.e., $\langle gnd | \beta(t) \rangle = 0$ for all $t$.
By substituting this anzatz and the new effective Hamiltonian from Eq.~(\ref{eq:H_eff_cascade}) into the unnormalized evolution equation Eq.~(\ref{eq:unnormalized_dpsi}) and projecting onto the subspace where the photon qubit is in $|g\rangle$, we are able to compute the equation of motion for the excited state $|\beta(t)\rangle$:
\begin{multline}\label{eq:dbeta}
    \frac{d}{dt} |\beta(t) \rangle = \Big( - i H_\text{sys} - \tfrac{1}{2} \sum_{j = 0} L_j^\dag L_j \Big) |\beta(t) \rangle - \xi(t) L_0^\dag\, |gnd \rangle .        
\end{multline}

This equation takes the satisfying form in which $|\beta(t)\rangle$ has a homogeneous solution that evolves under the system effective Hamiltonian $H_\text{sys} - \tfrac{1}{2} \sum_{j = 0} L_j^\dag L_j$, as well as a particular solution that is given by the source term $\xi(t) L_0^\dag\, |gnd \rangle$.  The source term is proportional to both the incoming wave-packet $\xi(t)$ and the creation of a single excitation by the adjoint of the channel 0 emission operator, $L_0^\dag$.   This is a key result, allowing the description of absorption of a single photon in a single quantum trajectory. The absorption of the single photon creates a distributed `bright state', which is a single excitation that may be in superposition across many chromophores in the sample, with the details of this superposition being dependent on the distribution of dipole operators contributing to $L_0$.  For an atomic ensemble distributed in space, this excitation is sometimes referred to as a `spin wave'~\cite{hammerer_quantum_2010}. For an ideal ensemble of identical atoms located at the same position in space, this superposition is the well known Dicke state, i.e., the $N = 1$ superradiant excitation~\cite{dicke_coherence_1954}.   

In the case of PSII, all chromophore sites are approximately co-located due to its small size ( $\lesssim 30$ nm~\cite{bennett_structure-based_2013}) relative to the relevant optical wavelength (680 nm). However the chromophore transition dipoles are not aligned in PSII and possess a rich and complex structure.  Because the $L_0^\dag$ operator is proportional to the input polarization dotted into the collective dipole operator, $L_0^\dag \propto \mathbf{e}_0 \cdot \mathbf{D}^\dag$, we see that by varying the input polarization, we will excite different chromophores to differing degrees.  Analysis of the chlorophyll transition dipoles for PSII shows that these are neither fully aligned nor do they constitute a purely disordered system. In the following we analyze the collective properties of the chromophore transition dipoles in more detail.

\subsection{Bright states and dipole couplings \label{sec:bright_states} }
\begin{figure*}
\includegraphics[width=1.0\textwidth]{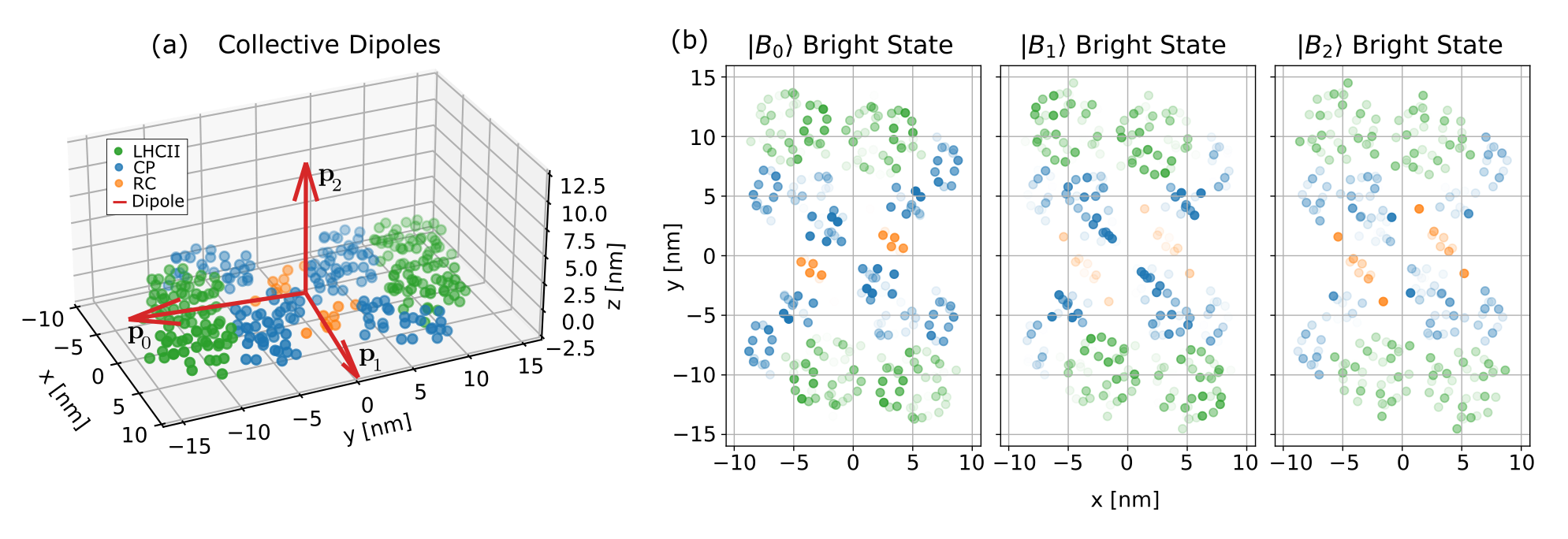}
\caption{\label{fig:collective} Collective dipole orientations: (a) The direction and relative magnitude of the collective dipoles $\mathbf{p}_0 - \mathbf{p}_2$. Superimposed are the center positions of each contributing dipole, color coded to their sub-complex designation: LHCII (green), CP (blue), and RC (orange).  (b) Bright states $|B_0\rangle - |B_2\rangle$ are shown as the brightness in each site, $\propto |\mathbf{v}_i \cdot \mathbf{d}_n |^2$, normalized to the maximum over all sites and polarizations.}
\end{figure*}

Before considering the general scenario, first consider two orthogonal input polarizations $\mathbf{e}_a$ and $\mathbf{e}_b$, with $\mathbf{e}_a \cdot \mathbf{e}_b = 0$. These two polarization states will generate two generally distinct (unnormalized) bright states, $|B_a\rangle$, $|B_b\rangle$, where
\begin{equation} \label{eq:bright}
    |B_\alpha\rangle  = \mathbf{e}_\alpha \cdot \mathbf{D}^\dag |gnd\rangle = \sum_{n = 1}^N \mathbf{e}_\alpha \cdot\mathbf{d}_n\, |n\rangle
\end{equation}
for $\alpha \in \{ a, b \}$, with $|n\rangle \equiv \sigma_+^{(n)} |gnd \rangle$ the state in which only the $n^{th}$ chromophore is excited.  Now although the input polarization vectors are orthogonal, the bright states may or may not have significant overlap. In fact, if all $N$ dipoles were to be aligned at 45$^\circ$ between $\mathbf{e}_a$ and $\mathbf{e}_b$, then $|B_a\rangle = |B_b \rangle $.  However, there does in fact exists a natural set of coordinates $\{\mathbf{v}_i,\, i = 0,1,2\}$ such that the corresponding bright states $|B_{i}\rangle$ are either orthogonal or zero.  We will now find such a basis. 
Assuming that this basis exists, we have the relationship,
\begin{equation}
\begin{split}
    \langle B_i | B_j \rangle = \| B_i\|^2\, \delta_{ij}  &= \sum_{n = 1}^{n} (\mathbf{v}_{i} \cdot \mathbf{d}_n) ( \mathbf{v}_j \cdot \mathbf{d}_n ) \\
    &= \mathbf{v}_i\cdot \left(\sum_{n = 1}^N \mathbf{d}_n \mathbf{d}_n\right)\cdot \mathbf{v}_j
\end{split}
\end{equation}
where $\mathbf{d}_n \mathbf{d}_n$ is the dyadic (outer) product of the two vectors.  In other words, $\{ \mathbf{v}_i \}$ is a set of orthonormal eigenvectors of the $3\times 3 $ matrix  $\boldsymbol{\mathsf{M}} \equiv \sum_{n = 1}^N \mathbf{d}_n \mathbf{d}_n$ with eigenvalues $\|B_i\|^2$.  Furthermore, the vectors
\begin{equation}
    \mathbf{p}_i = \|B_i\|\, \mathbf{v}_i, 
\end{equation}
are the collective transition dipoles for these orthogonal bright states.  Diagonalizing the matrix $\boldsymbol{\mathsf{M}}$ for the 324 chromophores of PSII yields the directions and scaled magnitudes of the vectors $\mathbf{p}_i$ that are shown in Fig. \ref{fig:collective} (a), together with a three-dimensional scatter plot of the center positions of the individual dipoles. It is evident that rather than displaying a random or arbitrary set of orientations, the collective dipole moments are actually well aligned with the natural geometry of the total complex.  In terms of absolute units, the collective dipoles have magnitudes  $\|\mathbf{p}_0\| = 44.33 \text{ Debye}$, $\|\mathbf{p}_1\| = 41.29 \text{ Debye}$, and $\|\mathbf{p}_3\| = 35.54 \text{ Debye}$.  Note that $\mathbf{v}_0$ is the input polarization that gives the largest collective dipole moment.  

Fig. \ref{fig:collective} (b) shows the relative contributions of the individual dipoles to each bright state.  For each of the three bright states, we denote the X-Y position of the individual dipoles with a point with variable color and brightness. The color of each point corresponds to the subcomplex type, while the brightness is proportional to $|\mathbf{v}_i \cdot \mathbf{d}_n|^2$, normalized to the maximum over all polarizations and positions.  A striking feature of these distributions is that the reaction center is seen to exhibit a strong polarization dependence, with the primary D1 and D2 chlorophyll molecules as well as two strongly coupled molecules (Chl$_\text{D1}$ and Chl$_\text{D2}$) preferentially coupling to $\mathbf{v}_0$ polarization.

\section{Results - No Phonons \label{sec:simulations} }

We now present numerical simulations of the SSE for a single PSII complex, with the goal of understanding both the photon averaged evolution as well as the correlations between the observed field channels (corresponding to transmission and fluorescence) and the system dynamics that are revealed by the individual trajectories. The key distinction will be between trajectories where an exiting photon is observed, and trajectories for which no exiting photon is observed.  This section considers the limit of PSII in isolation, uncoupled to a phononic environment.  Sec.~\ref{sec:phonons} contrasts this picture with a model in which PSII is coupled to a Markovian phonon bath.

We assume that all detectors have a collection solid angle $\Delta \Omega \approx 1.03 \text{ steradians}$, resulting in an efficiency factor $\eta = 0.123$.  We fix the pulse envelope to have a Gaussian profile with standard deviation in time of $\sigma = 0.1 \text{ ps}$, which corresponds to a frequency bandwidth of $\sigma^{-1} = 333.56\text{ cm}^{-1}$.  The peak arrival time of the pulse is set to be $8 \sigma$ after the initial time $t_0$;  $t_{pk} = t_0 + 8 \sigma$.    These parameters of the incident photon and optical setup are within the experimental capabilities of current quantum optical technology. 

We choose the incident polarization and center wavelength of the wave packet $\xi(t)$ to approximately maximize the probability of photon absorption by the PSII complex.  This is done by first aligning the incoming polarization, $\mathbf{e}_y$, to maximally couple to the bright state with the largest collective dipole moment $|B_0\rangle$.  This has the effect of slightly rotating our coordinate system so that $\mathbf{e}_y  = - \mathbf{v}_0$, $\mathbf{e}_x = \mathbf{v}_1$ and $\mathbf{e}_{z} = \mathbf{v}_2$. We then set the central carrier wavenumber $k_0$ to be resonant with the specific energy eigenstate that has the largest overlap with the bright state $|B_0\rangle$.  Explicitly, given that $E_m$ ($|E_m\rangle$) are the eigenvalues (eigenvectors) for the system Hamiltonian $H_{sys}$, we set $k_0 = E_{m^\star}/ \hbar c$, where $E_{m^\star} = \max_{\{E_m\}} |\langle E_m | B_0 \rangle |^2$.  For our model of PSII with no phonon coupling and hence no reorganizational energy, this results in $k_0 = 2\pi \times  14,955 \text{ cm}^{-1}$, i.e., a vacuum  wavelength $\lambda_0 = 668.66 \text{ nm}$.

To implement the simulations we use the two step continuous time sampling procedure outlined in \cite{dum_monte_1992}. For each trajectory we generate a jump time, $\tau$, and then, given that a jump occurred at that time, we sample from a conditional distribution to determine in which channel the jump occurred.  Specifically, we use an inverse cumulative distribution function (CDF) sampling procedure where we first sample a value $u$ from the uniform probability distribution $\mathsf{U}_{[0, 1]}$ and then numerically solve for the time $\tau$ that satisfies the condition
\begin{equation} \label{eq:sample1}
    u = \|\tilde{\psi}_{tot}(\tau)\|^2. 
\end{equation}
This can be seen to be equivalent to inverse CDF sampling of the jump probability by the following argument.  The probability to observe zero counts across all channels in the interval $[t_0, \tau]$ is known as the waiting time distribution and is given here by $\|\tilde{\psi}_{tot}(\tau)\|^2$,  with $\tilde{\psi}_{tot}$ defined by Eq.~(\ref{eq:unnorm_psi_tot}).  $\|\tilde{\psi}_{tot}\|^2$ is the probability to have either an excitation in the incoming field (source qubit) or an excitation in the PSII system, i.e., it is the total probability for the incident excitation to be ``upstream'' of the detectors.  Furthermore, the total system contains only a single excitation in the system, so in any given trajectory there can only be either 0 or 1 counts observed, never 2 or more.  Thus the cumulative probability to see 1 count in the interval $[t_0, \tau]$ is the complement of the waiting time, $1 - \|\tilde{\psi}_{tot}(\tau)\|^2$.   Seeking a time $\tau$ when the probability for zero counts crosses a uniformly random threshold value $u$ from above is then equivalent to searching for the time $\tau$ at which the probability to see 1 count crosses the uniformly random threshold value $u' = 1 - u$ from below.

Given that a jump occurs at the jump time $\tau$, we must also determine into which jump channel $j$ the output photon is emitted. The unconditional probability to see a jump in any channel in time interval $[t, t + \Delta t]$ is $\sum_j \mathsf{prob}_j(t) \ll 1$,  where $\mathsf{prob}_j(t)$ is given in Eq.~(\ref{eq:prob_j}).  However the re-normalized conditional probability for observing a jump in channel $j$, $\mathsf{prob}_j|_\tau$, must sum to 1.  Therefore, our second sampling step is to sample from following the discrete conditional distribution, 
\begin{equation}
\begin{split}
    \mathsf{prob}_j|_\tau &\equiv \mathsf{prob}_j(\tau_-)/ \sum_k \mathsf{prob}_k(\tau_-) \\
    &= \langle L_j^\dag L_j \rangle_{\tilde{\psi}_{tot}(\tau_-)} / \sum_{k} \langle L_k^\dag L_k \rangle_{\tilde{\psi}_{tot}(\tau_-)}.
\end{split}
\end{equation}
Here expectation values of the operators $L_j^\dag L_j$ are taken with respect to the unnormalized state $\tilde{\psi}_{tot}$ evaluated at the left limiting time $\tau_-$, i.e., just before the jump at time $\tau$. 

\begin{figure}
\includegraphics[width=1.0\columnwidth]{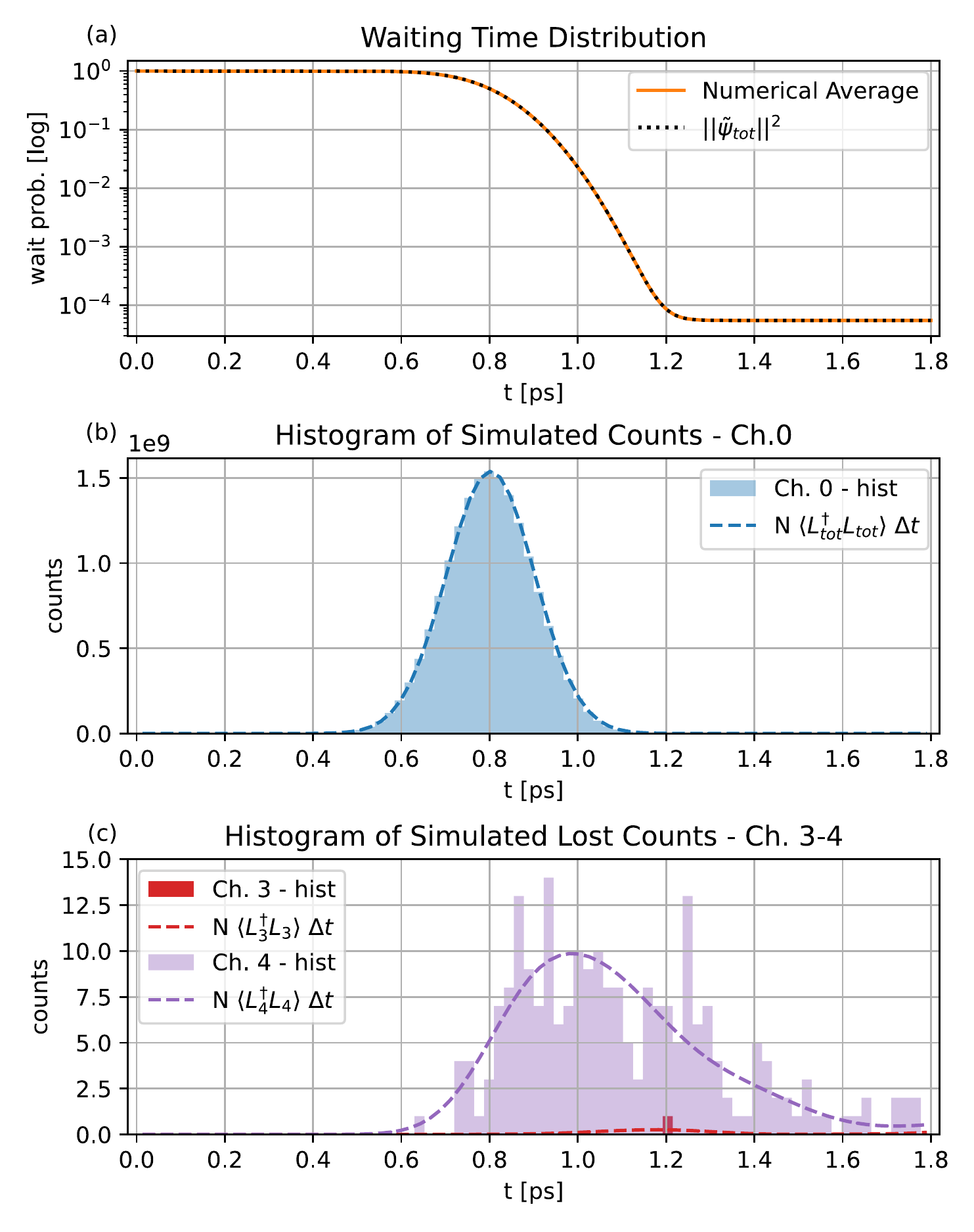}
\caption{\label{fig:counts} Waiting time distribution and photon count histograms for PSII fluorescence following absorption of a single photon, obtained from $N = 2^{34} \approx 10^{10}$ sample trajectories. (a) Probability distribution for having to wait a time $t$ to observe the first jump (waiting time distribution).  (b) Histogram of simulated photon emission jumps observed in the transmitted detector, channel 0.  (c) Histogram of simulated photon counts observed in the non-transmitted channels.   Also shown here are the expected number of counts for two of these channels; $N \langle L_{i}^\dag L_{i} \rangle \Delta t, \,\, i=3,4$, with $\Delta t = 22.5 \text{ fs}$ (see Fig.~\ref{fig:schematic}). }
\end{figure}

Fig. \ref{fig:counts} shows the waiting time distribution for our model of the chromophores in PSII (panel (a)), together with a histogram of the observed counts in panel (b). These results were obtained with $N = 2^{34} \approx 10^{10}$ sample trajectories.   The four orders-of-magnitude decrease seen in the waiting time distribution over the time interval 0.6 - 1.2 ps is due to the fact that the $w(t)$ term in $\tilde{\psi}_{tot}$ tends to zero after arrival of the pulse, while the corresponding system excitation only reaches a final value of $5.505 \times 10^{-5}$. This is consistent with the characteristic excitation probability of chlorophyll chromophores under single photon conditions~\cite{chan_single-photon_2018}. We note that the equally spaced grid of time points used to numerical integrate Eq.~\ref{eq:unnorm_psi_tot} and then to obtain the waiting time distribution, does not result in an equally spaced grid of waiting time probabilities.  In order to prevent any bias this may introduce, we linearly interpolate between neighboring waiting time probabilities, which allows for simulated jump times that fall outside of the initial time grid. 

Of the total number of $\sim 10^{10}$ trajectories, 945,664 resulted in no outgoing photons, i.e., the sampled jump time would be greater than the final simulation time $t_f = 1.8 \text{ ps}$.  Nearly all of the remaining trajectories resulted in a jump in channel 0, which is the channel that responds in the case of pure transmission.  228 trajectories resulted in photon detection in the environmental channels 3 and 4. Of these trajectories, 227 resulted in photon detection in channel 4, representing a photon lost to the exterior environment with the same bright state coupling ($|B_0\rangle$) as the incident photon, and therefore also $\mathbf{e}_y$ polarized. Only 1 trajectory resulted in photon loss to the environment in channel 3, corresponding to coupling to the bright state $|B_1\rangle$ and a lost $\mathbf{e}_x$ polarized photon. Fig. \ref{fig:counts} (c) shows the histogram and expected count rates for these two environmental channels.  No trajectories resulted in counts in the orthogonal observation channels 1 and 2, or in the third environmental loss mode, channel 5. 
%[data saved from May 18, 2021.]

This two step sampling procedure makes generating single photon trajectories extremely efficient, particularly when working on a fixed time interval $[t_0, t_f]$.  The efficiency is gained by realizing that for each trajectory either the jump has not yet occurred, and the system has deterministically followed the smooth evolution of Eq. (\ref{eq:unnormalized_dpsi}), or a jump has occurred and the total system has been projected into the zero excitation state $|g\rangle|gnd\rangle$.  Therefore the problem of producing single trajectories reduces to the problem of sampling the jump times.  In fact, because every decay channel $j$ ends in the same ground state, the second sampling step has no effect on the given trajectory.  Only the fact that a photon was detected matters to the posterior state, not where the photon was detected.  These simulations were written in Python 3 and made use of the standard NumPy libraries. They were executed on a single Intel core i7 CPU with 16 GB of RAM.  The $2^{34}$ samples of Fig~\ref{fig:counts} required $\sim70$ hours of computation time. 

The only piece of information that matters to the $m^{th}$ trajectory is the random jump time $\tau_m$.  Each trajectory can then be written in terms of the renormalized, `no jump evolution' state vector and the post jump ground state, with the use of two indicator functions $\chi_{[t_0, \tau_m)}(t)$ and $\chi_{[\tau_m, \infty)}(t)$.  The wave function conditioned on the jump is then given by 
\begin{equation} \label{eq:nonoverlappingsupport}
    |\psi_m(t)\rangle = \chi_{[t_0, \tau_m)}(t) \frac{1}{\|\tilde{\psi}_{tot} \| } |\tilde{\psi}_{tot}(t)\rangle + \chi_{[\tau_m, \infty)}(t) |g, gnd\rangle. 
\end{equation}
This expression is very useful for computing the density matrix by averaging over trajectories because when two indicator functions are referred to non-overlapping intervals as in Eq.~(\ref{eq:nonoverlappingsupport}), they have disjoint support and thus there is no possible interference between the two terms of $|\psi_m(t)\rangle$.  Therefore the ensemble average over the trajectories is equal to
\begin{equation}
\begin{split}
    \rho(t) = \frac{1}{N}\sum_{m = 1}^N & \chi_{[t_0, \tau_m)}(t)\, \frac{1}{\|\tilde{\psi}_{tot} \|^2 }|\tilde{\psi}_{tot}(t)\rangle \langle \tilde{\psi}_{tot}(t)| \\
    &+ \chi_{[\tau_m, \infty)}(t)\, |g, gnd \rangle \langle g, gnd |.
\end{split}
\end{equation}
The only quantities that depend upon the jump time here are the indicator functions, so that computing the density matrix reduces to making an average over a function that is equal to one before and zero after the jump time, and conversely over its complement. Furthermore, the sum over all indicator functions $\sum_{m } \chi_{[t_0, \tau_m)}(t)$ counts the number of trajectories with a jump time $\tau_m > t$.  Therefore the sample average is an estimate for the probability to have not seen any photons in the time interval $[t_0, t]$, i.e., the waiting time distribution.  We then have the convergence 
\begin{equation}
    \lim_{N \rightarrow \infty} \frac{1}{N}\sum_{m = 1}^N  \chi_{[t_0, \tau_m)}(t) = \| \tilde{\psi}_{tot}(t)\|^2. 
\end{equation}
The numerical test of this convergence is demonstrated in Fig. \ref{fig:counts} (a) where we plot the numerical average together with the integrated value of the waiting time distribution, $\| \tilde{\psi}_{tot}(t)\|^2$.  

This leads to the following intuitive form of the density matrix 
\begin{equation}\label{eq:rho}
    \rho(t) = |\tilde{\psi}_{tot}(t)\rangle \langle \tilde{\psi}_{tot}(t)| 
    + (1 - \|\tilde{\psi}_{tot}\|^2) \, |g, gnd \rangle \langle g, gnd |.
\end{equation}
This expression is simply the realization of the long standing notion in quantum optics that an open quantum system evolves under a decaying effective Hamiltonian, represented by the first term, with the addition of trace preserving `feeding' terms which preserve the norm of the state, represented by the last term. Here we are fortunate to have a simple model in which every scattered photon returns the system to an identical ground state. In this situation the single feeding term is guaranteed to take this particularly simple form.

\subsection{Trajectories and Bayesian interpretation}\label{sec:Bayes}
\begin{figure}
\includegraphics[width=1.0\columnwidth]{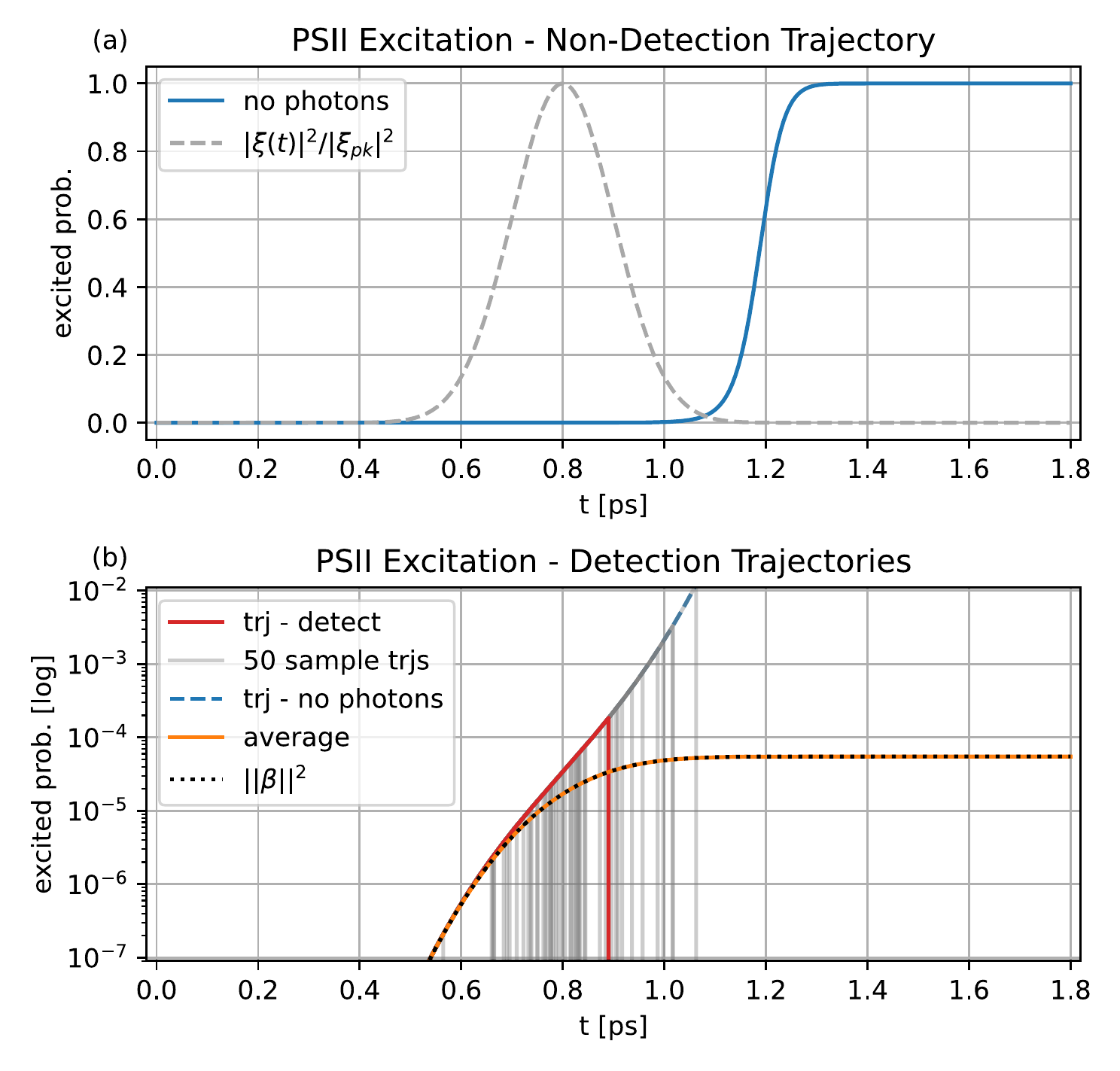}
\caption{\label{fig:cond_0_vs_cond_1} Excited state trajectories showing the total probability for the molecule to be excited as a function of time for different measurement realizations.  (a) Excitation probability of a single trajectory for which no photons were emitted in any direction (solid blue line). The scaled incoming photon pulse is also shown (dashed grey line).  (b) 50 sample trajectories in all of which a photon is detected in channel 0 are shown (grey), with one highlighted trajectory (solid red line). Each trajectory follows the same non-detection curve (dashed blue line) until the detection time, when the quantum jump causes all excited state probabilities to transition sharply to zero. Also shown is the average excitation probability taken over a total of $N = 1 \times 10^8$ sample trajectories (solid orange line). Of these, 5413 trials resulted in no measured photons, yielding a final average excitation probability $\sim 5 \times 10^{-5}$.  This average matches the analytical expected average $\|\beta\|^2$ (dotted black line).}
%[saved data from April 23, 2021.] }
\end{figure}

Fig. \ref{fig:cond_0_vs_cond_1} (a) shows the deterministic evolution of an individual trajectory that is conditioned upon not observing any outgoing photons in any outgoing mode.  This complete information of no photons means that the system always follows the smooth deterministic equation of Eq. (\ref{eq:unnormalized_dpsi}), renormalized at every time step.  Here the lack of an exiting photon provides a clear indication that the photosynthetic system must have absorbed the photon, leading to a dramatic rise in the probability of being in an excited state.

Fig. \ref{fig:cond_0_vs_cond_1} (b) shows 50 typical trajectories, with one highlighted in red to show the characteristic sharp jump when the photon is emitted.  Each trajectory follows the same deterministic no-jump evolution (dashed blue line) up to the jump time $\tau$.   For the red highlighted trajectory, this jump occurred at $t \sim 0.9 \text{ ps}$. After this time the molecular complex is known with certainty to be in its ground state with zero excitation probability, i.e.,  for times $t \ge \tau$, the probability to be in the excited state is zero.  This panel also shows the average probability of excitation over time (solid orange line), computed over all $N = 1 \times 10^8$ trajectories that were simulated.  This average makes a smooth transition from 0 to a value of $\sim 5\times 10 ^{-5}$, within a time window that corresponds with the arrival of the incoming wave packet.  This numerical average also confirms the analytic prediction of Eq.~(\ref{eq:rho}) (dotted black line).  The average excitation probability is consistent with expectations for single photon excitation of chlorophyll molecules in smaller complexes~\cite{chan_single-photon_2018}. 

The conditional probability of excitation demonstrated by these measurement-based quantum trajectories can be interpreted in terms of a Bayesian probability for an excitation to be either in the incoming field or in absorbing system.  To see this, note that the probability to not see any jumps is given by the norm squared of the unnormalized total state, $\|\tilde{\psi}_{tot}(t)\|^2$, Eq. \ref{eq:unnorm_psi_tot}, which is equal to 
\begin{equation} \label{eq:norm_sq}
    \|\tilde{\psi}_{tot}(t)\|^2 = w(t) + \| \beta(t) \|^2.
\end{equation}
Given that no jumps have occurred, the conditional probability for the system to be in the 1-excitation manifold is given by 
\begin{equation}
    \mathsf{prob}_{ex}(t) = \langle \tilde{\psi}_{tot} | \Pi_{ex} | \tilde{\psi}_{tot} \rangle / \|\tilde{\psi}_{tot}\|^2,
\end{equation}
where $\Pi_{ex}$ is the projector onto the 1-excitation manifold.  Noting that $|\beta \rangle$ ($|gnd \rangle$ ) is in the 1 (0) excitation subspace, the conditional probability then reduces to
\begin{equation}
    \mathsf{prob}_{ex}(t) = \frac{\|\beta(t)\|^2}{w(t) + \| \beta(t) \|^2}.
\end{equation}
Conversely, the conditional probability for the atomic system to be in the zero excitation subspace, i.e., in the state $|gnd \rangle$ is
\begin{equation}
    \mathsf{prob}_{gnd}(t) = \frac{w(t)}{w(t) + \| \beta(t) \|^2}.
\end{equation}
Thus an observation that no jumps have occurred implies that there are only two remaining paths for an excitation conserving system, which are the posterior Bayesian probabilities.  
Initially, the system is in its ground state $|gnd\rangle$ and the excitation is in the incoming pulse with $w(t) = 1$, $\|\beta (t)\|^2 = 0$.  After the pulse passes, $w(t) \rightarrow 0$ and so $\mathsf{prob}_{ex}(t) \rightarrow 1$, even if the average absorption is small, as measured when $\|\beta \|^2 \ll 1$. The crossover point of $\mathsf{prob}_{ex}(t) = \mathsf{prob}_{gnd}(t) = 1/2$ occurs when $w(t) = \|\beta(t)\|^2$.  This equality is satisfied at later times for weaker overall absorption probability.  This can be seen by comparing the crossover time for PSII in Fig. \ref{fig:cond_0_vs_cond_1} (a) to the corresponding time for a pentameric chromophore system in Fig. \ref{fig:coherent_vs_single} (a) below, for which the chromophore-field couplings were artificially increased by a factor of $\sim 5\times 10^3$ to a value of $1/( 5\text{ ps} )$ (see below).

\subsection{Comparison with Conditional Excitation from a Coherent State Source}

Here we compare the conditional evolution under excitation pulse with a single photon with the corresponding evolution under an excitation pulse of a weak coherent state.  We make this comparison with a highly idealized model, in order to emphasize the role of the photon detection measurements and the quantum statistical effects. Our model for this comparison involves only 5 chromophores interacting with a one dimensional mode.  A pulse of laser light has an indefinite number of photons, which when modeled by a coherent state exhibits photon number statistics that follow a Poisson distribution.  It is then easy to verify that a coherent state with a mean photon number $\bar{n} = 1$ has equal probability $p_0 = p_1 = e^{-1} \sim 0.37$ of containing 0 or 1 photons.  

However, the remaining $\sim 0.26$ probability represents pulses with multiple photons. So a coherent state has the possibility of creating multiple excitations, in addition to creating just a single excitation.  In order to accommodate multiple excitations we must utilize a much larger system Hilbert space, which is the reason that our comparison here is limited to a simple 5 chromophore model system rather than to the full PSII.

The uncertainty in photon number leads to a strong reduction in the amount of information conferred by not measuring a transmitted photon. In the case of a single photon input, not measuring a transmitted photon guarantees that the photon had to be absorbed.  However, for a coherent state, not detecting a photon means either all the photons in the pulse were absorbed or the pulse initially contained zero photons.  We show below that this ambiguity leads to a strong reversion of the conditional excitation probability to the unconditional average, which is quite different from the conditional excitation probability from a single photon pulse.   

\begin{figure}
\includegraphics[width=3in]{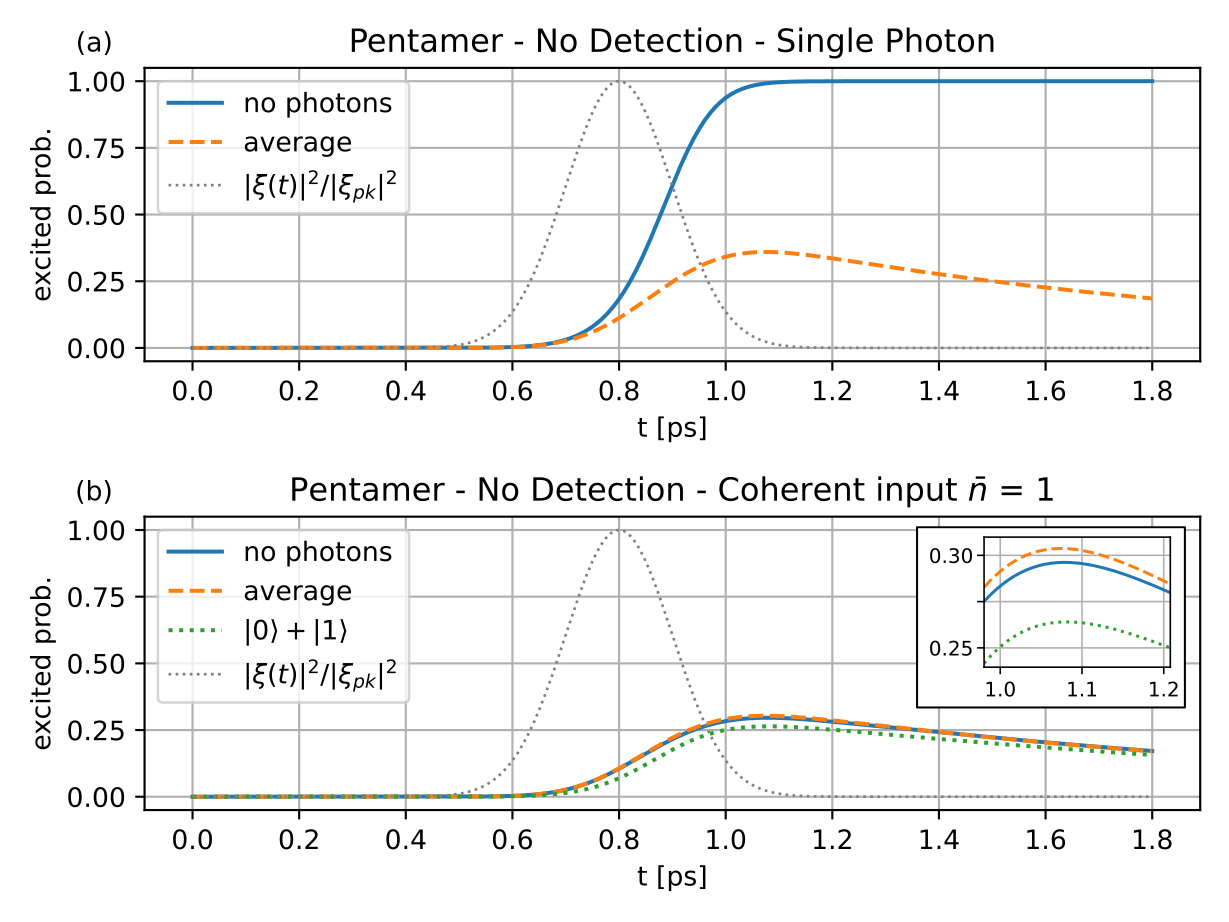}
\caption{\label{fig:coherent_vs_single} Single photon and coherent state conditional excitation comparison for a pentamer containing five Chl chromophores, all aligned and identically coupled to the field with coupling rate $\gamma_0 = 1/(5 ps)$. (a) The single photon input excitation probability, conditional upon not detecting an output photon (blue line) and the expected average (dashed orange line).  (b) The excitation probability is shown for a  coherent state input with mean photon number equal to 1, conditioned upon non-detection (blue line) and the average master equation (dashed orange line).  The dotted green line shows the excitation probability for an equal superposition between 0 and 1 photons, conditioned upon non-detection of a photon.  The peak values of all three curves are shown in the inset.}
\end{figure}

Fig. \ref{fig:coherent_vs_single} shows the comparison for the pentamer model between conditional excitation from a single photon pulse and from a coherent state with mean photon number equal to 1.   The single photon conditional evolution is generated via Eq.~(\ref{eq:dbeta}) and the average is given by Eq.~(\ref{eq:rho}). The conditional evolution for a input coherent state is well known in quantum optics ~\cite{wiseman_quantum_2010} and in general this follows the non-linear stochastic differential equation (Ito form): 
\begin{equation}
\begin{split}
    d | \psi_{ch}(t) \rangle &= \Big(\tfrac{1}{2} \langle L^\dag L \rangle_{\psi_{ch}(t)} + \tfrac{1}{2}  \left\langle \xi(t) L^\dag + \xi^*(t) L \right\rangle_{\psi_{ch}(t)} \\
    &\qquad - \tfrac{1}{2} L^\dag L - \xi(t) L^\dag - i H  \Big) | \psi_{ch}(t) \rangle\, dt \\
    &\ + \left( \frac{L + \xi(t)}{ \big\| (L + \xi(t)) \, | \psi_{ch}(t) \rangle \big\| }  -  1 \right)\, | \psi_{ch}(t) \rangle\, d N(t).
\end{split}
\end{equation}
Here $dN(t)$ is a pure jump counting increment, i.e., the measurement outcome for time increment $dt$, which is equal to zero for all time in a realization with no observed photon counts.  Thus for the zero counts trajectory, the final term is always zero and the equation is no longer stochastic. 

In terms of the density matrix, a coherent laser simply induces a local Hamiltonian rotation upon the system and so the average evolution for the coherent state input is easily calculated with a standard density matrix master equation~\cite{chan_single-photon_2018}: 
\begin{equation}
\begin{split}
    \frac{d}{dt}\rho_{ch}(t) &= -i [H, \rho_{ch}(t)] - i [-i\xi(t) L^\dag + i \xi^*(t) L, \rho_{ch}(t) ] \\
  &\quad + L \rho_{ch} L^\dag - \tfrac{1}{2} L^\dag L \rho_{ch} - \tfrac{1}{2} \rho_{ch} L^\dag L.
\end{split}\end{equation}

We note that this simplified one-dimensional model eliminates any coupling to unobserved radiation modes and assumes perfect input to output coupling.  Furthermore, it assumes that all sites are identically coupled with the same coupling rate $\gamma_0 = 1/(5 \text{ ps})$ (which is considerably higher than the values for PSII) and that the local Hamiltonian $H = 0$.  The incident pulse has a Gaussian pulse shape with a standard deviation $\sigma = 0.1\text{ ps}$.   The artificially high coupling rate was chosen so that the peak unconditional absorption would be $O(1)$, rather than the $\sim 10^{-5}$ seen in PSII~\cite{chan_single-photon_2018}, and thereby ease the comparison to the conditional dynamics. None of these simplifications affect the comparison between single photon and coherent state pulses. 

Fig. \ref{fig:coherent_vs_single} (a) shows numerical integration of a single photon input with the characteristic rise in the excitation probability conditional upon not detecting a transmitted photon.  It also shows the average evolution, which has a much larger peak value due to the artificially large coupling rate in this demonstration.  Fig. \ref{fig:coherent_vs_single} (b) shows the corresponding conditional excitation probability for an incident coherent state with mean photon number $\bar{n} = 1$. Here the conditional excitation is nearly identical to that of the unconditional excitation probability, confirming the prediction made above.

The effect of the vacuum component of the coherent state input can be understood analytically by first considering a slightly simpler setting, namely an input state that is an equal superposition of 0 and 1 photons.  In this situation the cascaded source model of Sec. \ref{sec:SSE} can still be utilized, however a different source model would be needed to construct a superposition that included photon numbers greater than 1.  Thus the cascaded initial state would now be $|\tilde{\psi}_{tot}(t_0)\rangle = \frac{1}{\sqrt{2}}(|g\rangle + |e\rangle)\otimes|gnd\rangle$.  The zero excitation part of the total superposition, $|g\rangle|gnd\rangle$ is stationary under the joint evolution,  while the excited part evolves identically to the pure single photon input. Therefore the total unnormalized state will always be of the form
\begin{equation}
    |\tilde{\psi}_{tot}(t)\rangle = \tfrac{1}{\sqrt{2}} |g\rangle|gnd\rangle + \tfrac{1}{\sqrt{2}} \left(\sqrt{ w(t)} |e\rangle |gnd\rangle + |g\rangle |\beta(t) \rangle \right) 
\end{equation}
where $|\beta(t) \rangle $ still evolves under Eq.~(\ref{eq:dbeta}). Reproducing the Bayesian arguments of the previous section results in a probability of excitation
\begin{equation}
    \mathsf{prob}_{ex}(t) = \frac{\tfrac{1}{2}\|\beta(t)\|^2}{\tfrac{1}{2}\left(1 + w(t) + \| \beta(t) \|^2\right)}.
\end{equation}
This expression is also plotted in Fig. \ref{fig:coherent_vs_single} (b) (dotted green line).  The fact that $|\tilde{\psi}_{tot}(t)\|^2 \le \|\tilde{\psi}(t_0)\|^2 $ implies that $\|\beta(t)\|^2 \le 1$.  This means that $\mathsf{prob}_{ex}(t) \le 1/2$ for the input state $|0\rangle +|1\rangle$, with equality when $\|\beta(t)\| = 1$ and $w(t) = 0$.   

Returning to the comparison with a coherent state in Fig. \ref{fig:coherent_vs_single} (b), the inset shows that the probability of excitation conditioned upon observing no outgoing photons is greater for the coherent state than it is for the $|0\rangle +|1\rangle$ superposition. This slight increase is due to the fact that the coherent state has support over multiple photons and thus there is still some probability for multi-photon absorption.    

\subsection{Fluorescent rate detection \label{sec:fluorescence}}
In the previous sections we have shown that in experimental runs where no transmitted photon is detected following absorption of a single photon, the probability of the system to be excited is dramatically increased.  Here we consider what information about PSII is carried by the fluorescent photons, and in particular, by their emission times.  Specifically we calculate the average flux of our three detection channels (0, 1, and 2) for a longer time scale than the previous simulations.  Because the initial bright state $|B_0\rangle$ is not generally an energy eigenstate, we cannot expect $|\beta(t)\rangle$ to be stationary, even for times long after the photon wave-packet has interacted with the PSII chromophoric system.  Thus any intermediate dynamics between $t_{pk}$ and the emission time $\tau$ will show a signature in $|\beta(t)\rangle$.  

This can be seen by formally integrating the equation of motion for $|\beta(t) \rangle$, Eq.~(\ref{eq:dbeta}), which results in the integral expression
    \begin{equation}
        |\beta(t) \rangle = - \sqrt{\eta \gamma_0}\, \int_{t_0}^{t} ds\, e^{- i H_{eff} (t - s) }\,  \xi(s) |B_0\rangle.
    \end{equation}
It is natural to simplify the time dependence of this expression by working in the eigenbasis of $H_{eff}$.  However, this is complicated by the fact that $H_{eff}$ is non-Hermitian and fails to commute with its adjoint. Therefore its eigenvectors $\{ |z_n \rangle \}$ are not orthogonal, i.e., $\langle z_n | z_m \rangle \ne \delta_{nm}$.   Fortunately $H_{eff}$ is not a defective matrix and can still be diagonalized by a (non-unitary) similarity transformation.  

Choosing to work in the eigen-energy basis $\{|E_n\rangle \}$ for $H_{sys}$, we will utilize the transformation $Q \equiv \sum_n |z_n \rangle \langle E_n |$.  Note that if there were no decay and $H_{eff} = H_{sys}$ (see Eq.~(\ref{eq:H_eff_cascade})) then $Q$ is simply equal to the identity. Consequently for small decay relative to $H_{sys}$ we expect $Q$ to be close to the identity. Thus
\begin{equation}
    H_{eff} = Q\, \sum_{n} \zeta_n\, |E_n \rangle \langle E_n | \, Q^{-1},
\end{equation}
where $\{ \zeta_n \}$ are the complex eigenvalues.  Substituting this expression into  $|\beta(t)\rangle$ results in
\begin{multline}
    |\beta(t) \rangle = - \sqrt{\eta \gamma_0}\, \int_{t_0}^{t} ds\,Q \sum_{n = 1}^d e^{- i \zeta_n\, (t - s)\, |E_n \rangle \langle E_n |}\, Q^{-1}\\
     \times  \xi(s) |B_0\rangle.
\end{multline}
For times $t \gg \sigma$ (with $t_0 \ll \sigma$) and a Gaussian pulse, we can compute the integral with respect to $s$ analytically by taking the integration limits to $\pm \infty$, resulting in the approximate expression
\begin{multline}
    |\beta(t) \rangle \approx - (\sqrt{8 \pi}\, \eta \gamma_0 \sigma)^{1/2} \sum_{n = 1}^{d} e^{-i \zeta_n (t - t_{pk}) - \sigma^2 \zeta_n^2} |z_n\rangle  \\ \times \langle E_n|\, Q^{-1}\, |B_0 \rangle,
\end{multline}
where we have simplied the right hand side using the definition of $Q$.  This relation shows how the initial bright state $|B_0\rangle$ is mapped into the various eigenstates of $H_{eff}$ via the incident gaussian wave packet with temporal width $\sigma$.  

In order to effectively separate the transmitted photons from the fluorescence, we  consider only fluorescent counts with times $\tau \in [t_{cut}, t_f]$, employing an initial cut off time $t_{cut} = t_{pk} + 7 \sigma =  1.5 \text{ ps}$.  The rate of fluorescent counts detected in channels 0, 1,2, is 
    \begin{equation}
        R_i(t) \equiv \langle \tilde{\psi}_{tot}(t) | L_i^\dag L_i | \tilde{\psi}_{tot}(t) \rangle  = \eta \gamma_0\, |\langle B_i | \beta (t) \rangle |^2,
    \end{equation}
which follows from our definitions of $L_i$ in Eq.~(\ref{eq:measure_jump_ops}) and the bright states $|B_i\rangle$ in Eq.~(\ref{eq:bright}).   The overlap probability $|\langle B_i | \beta (t) \rangle |^2 $ contain terms $\propto e^{- i (\zeta_n - \zeta_m^*) t}$ and thus beat frequencies $\omega_{nm} \equiv \operatorname{Re} (\zeta_n - \zeta_m)$ will become evident as oscillations in the expected rate $R_i(t)$ and be revealed as distinct peaks in the Fourier spectrum of this.  Note that due to the relatively small contribution of the imaginary decay term, the real parts of the complex eigenvalues $\zeta_n$ are exceedingly close to the energy eigenvalues of the pure system Hamiltonian $H_{sys}$.  As a quantitative measure of this, the maximum of the minimum distance between the real part of $\zeta_n$ and the resonant frequencies of $H_{sys}$ is  
$\max_{n} \min_{m} | \mathsf{Re}(\zeta_n) - E_m | = 1.853 \times 10^{-5} \text{ cm}^{-1}$.

To resolve these beat frequencies we consider a significantly longer time scale with a final time, $t_f = 250 \text{ ps}$. In the actual system, this is likely to be well beyond the system coherence time, because of the environmental couplings to phonons that we have ignored. The simulations presented here are intended not to make quantitative predictions for a laboratory experiment, but rather to demonstrate the principle of the occurrence of system coherences in the rate of fluorescent photon detection. 

Since $R_i(t)$ is a rate with units of $1/\text{ps}$, its Fourier transform $\mathcal{F}[R_i](\omega)$ is unitless.  Also since the total excitation is only $\sim 10^{-5}$, it is useful to consider the unitless integrated photon flux (absorption probability) for the given time interval $[t_{cut}, t_f]$,
\begin{equation}
    \Phi_{tot} \equiv \sum_{i = 0}^2 \int_{t_{cut}}^{t_f}R_i(t)\, dt.
\end{equation}
For our parameter set, integrating over the time interval $[1.5 \text{ ps}, 250 \text{ ps}]$ gives an integrated photon flux of  $\Phi_{tot} = 2.42 \times 10^{-7}$. 

\begin{figure}
\includegraphics[width=1.0\columnwidth]{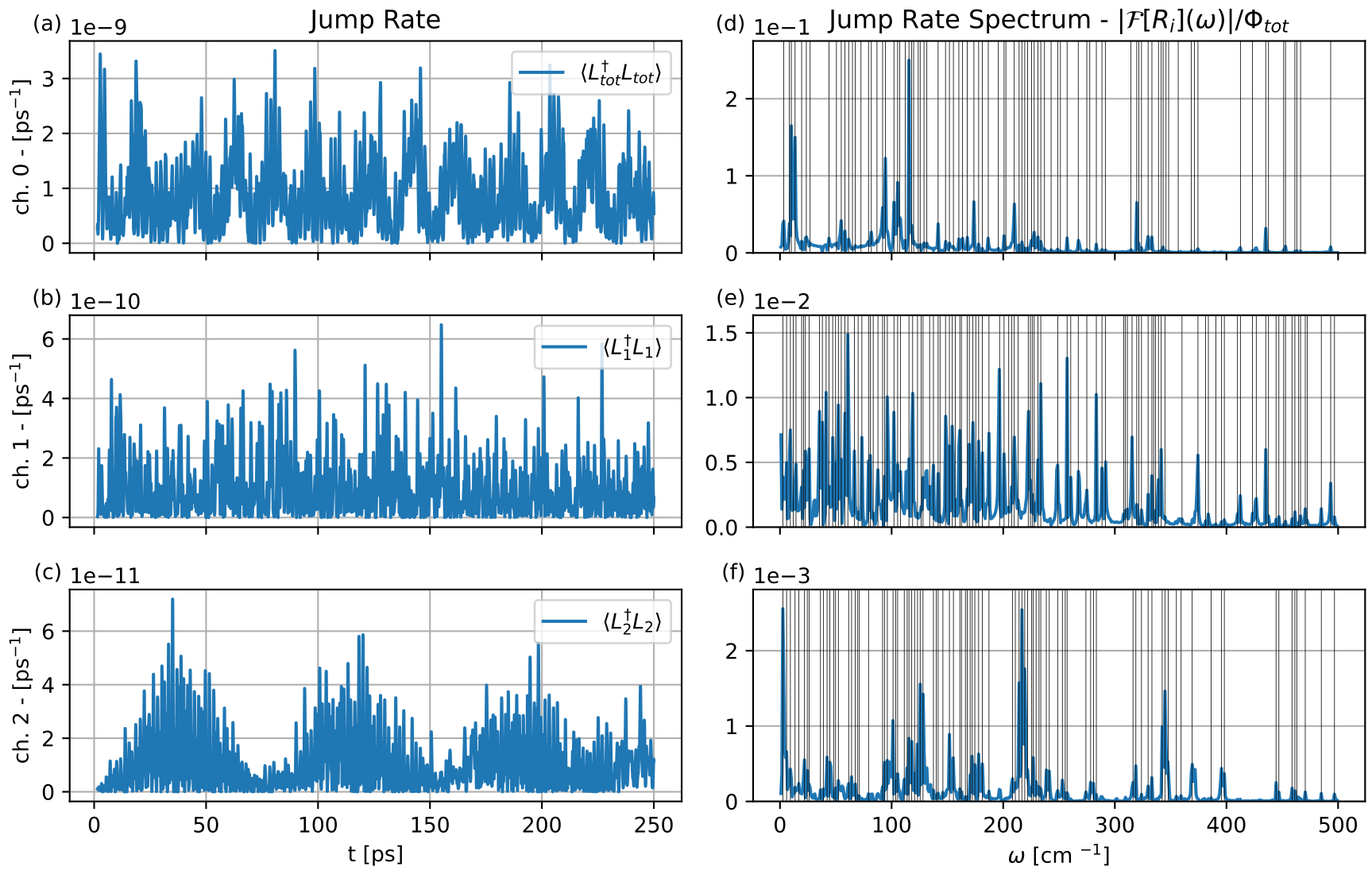}
\caption{\label{fig:spectra} Fluorescence spectra from PSII following absorption of a single photon, in the absence of exciton-vibration couplings.  Left column, panels (a-c):  Expected count rates for the 3 detectors in channels 0, 1, and 2.  Right column, panels (d-f):  Absolute frequency spectra of the count rates, normalized by total photon flux $\Phi_{tot}$ and excluding the zero frequency bin. Each spectral peak has a corresponding transition frequency $\omega_{nm} = \operatorname{Re}(\zeta_n - \zeta_m)/\hbar$, shown as a vertical line. }
\end{figure}

Figs. \ref{fig:spectra} (a-c) show the numerically integrated jump rates in the detection channels 0-2, respectively, for PSII. The corresponding absolute frequency spectra, $|\mathcal{F}[R_i](\omega) |$, normalized to $\Phi_{tot}$, are shown in Figs. \ref{fig:spectra} (d-f).  Each peak in this spectrum (we include all peaks having a maximum value at least $0.8\%$ of the largest peak in the spectrum), has a corresponding beat frequency. For each peak we have plotted the closest $\omega_{nm}$, vertical line.  The maximum deviation between a peak and the best fit frequency is less then half the minimum frequency spacing set by our maximum integration time, i.e., $|\omega_{pk} - \omega_{nm} | < \Delta \omega /2 = 0.422 \text{ cm}^{-1}$.  

\section{ Results - with phonons\label{sec:phonons} }
 
Here we consider what effect a simple model of the exciton-phonon interaction will have on the single photon absorption process.  Rather than the pure evolution under the non-Hermitian Hamiltonian, dissipation into the phonon environment now becomes a strong source of decoherence.  

One possible route is to consider a quantum trajectory unraveling of the emission/absorption of phonons by the system.  This will lead to a new type of jumps in a quantum trajectory, where in addition to the perfect detection of photons, the theory is also describing an ideal measurement of the phonon environment.  We will leave the microscopic understanding of the phonon statistics to another paper and instead restrict the scope of our observations in this work to the photons.  

Leaving the phonons as an unobserved degree of freedom requires the quantum trajectory to propagate an unnormalized conditional mixed state density operator, $\tilde{\rho}_c(t)$. \cite[section 7.4]{carmichael_open_1993}.  This corresponds to a partial unraveling of a general quantum master equation and cannot be described by an SSE.  The partial unraveling of a quantum master equation breaks the dynamics into periods of free evolution under a Liouvillian that lacks the ``feeding terms'' for the measured channels, i.e., the fluorescent photons, separated by measurement-induced jumps.  The feeding term is only applied when a photon is detected.  The waiting time between jumps is given by the trace of the unnormalized state.  In our setting we now have two classes of jump operators, namely the collective photon emission operators $L_i$, derived above, as well as the phonon dissipation operators $J_\alpha$, whose properties we will describe below.  

Here we will present a fully Markovian and (time-independent) Lindblad description of the exciton-phonon interaction.  Such a quantum master equation is equivalent to secular-Redfield theory \cite{breuer_theory_2007} and provides a simple albeit concrete starting place for modeling the exciton-phonon interaction. 

The full quantum master equation including both photonic and phononic dissipation channels defines the Liouvillian map
\begin{equation}\label{eq:phonons_unconditional}
    \frac{d\rho}{dt} = -i [H, \rho] + \sum_i \mathcal{D}[L_i] (\rho) + \sum_\alpha \mathcal{D}[J_\alpha] (\rho) \equiv \mathcal{L} (\rho) ,
\end{equation}
where the dissipator $\mathcal{D}[\cdot]$ is
\begin{equation}\label{eq:dissipator}
    \mathcal{D}[L](\rho) = L \rho L^\dag - \tfrac{1}{2} L^\dag L \rho - \tfrac{1}{2} \rho L^\dag L. 
\end{equation}
The first term of Eq. (\ref{eq:dissipator}) is known as the feeding term since it adds population to the state in such a way as to compensate for the population that is removed by the remaining terms.  

In the absence of any feeding terms, the state will evolve solely under the non-Hermitian Hamiltonian.  Thus between photon emissions, the unnormalized state $\tilde{\rho}_c(t)$ evolves under the conditional quantum master equation:
\begin{equation}
\begin{split} \label{eq:phonons_conditional}
        \frac{d\tilde{\rho}_c}{dt} =& -i H_{eff}\, \tilde{\rho}_c + i \tilde{\rho}_c H_{eff}^\dag + \sum_\alpha \mathcal{D}[J_\alpha] (\tilde{\rho}_c) \\
        =& \mathcal{L}( \tilde{\rho}_c) - \sum_i \mathcal{S}_i( \tilde{\rho}_c ),
\end{split}
\end{equation}
where the map defined by the feeding terms map is  $\mathcal{S}_i (\rho) \equiv L_i \rho L_i^\dag$ for all photon jump modes $i$.  When a photon is observed in mode $i$, $\tilde{\rho}_c$ undergoes the update $\tilde{\rho}_c \mapsto \mathcal{S}_i(\tilde{\rho}_c)$.  At any time $t$, the normalized conditional state $\rho_c(t)$ is then given by $\rho_c(t) = \tilde{\rho}_c(t) / \operatorname{Tr} \tilde{\rho}_c(t) $.

\subsection{The phonon model}
The exciton-phonon interaction is derived from a shifted oscillator model, where each exciton site is assumed to be coupled to an independent bath of oscillators whose equilibrium positions shift (see, e.g., ref.~\cite{ishizaki_quantum_2010} for a detailed derivation).  A time-independent Lindblad master equation is derived by constructing the system coupling operator in the energy basis $\{ |E_i\rangle \}$ and then applying second order perturbation theory with Markov and secular approximations.  Upon doing so, the phononic jump operators $J$, decompose into independent transitions between energy basis states, whose rates depend upon the originating excitonic site $n$, and the Bohr transition frequency $\omega_{ij} = E_i - E_j$ of the energy transition.  

Thus the index $\alpha$ in $\mathcal{L}(\rho)$ is a multi-index denoting a sum over both $n$ and all distinct $\omega_{ij}$ from the spectrum of $H_{\text{sys}}$.  The jump operators decompose into 3 distinct types, depending upon the value of the transition energy, i.e., 
\begin{equation}
    J_n(\omega_{ij}) =  \left\{ \begin{array}{cc}
        J^{(c)}_n(\omega_{ij})  & \omega_{ij} < 0 \\
        J^{(p)}_n & \omega_{ij} = 0 \\
        J^{(h)}_n(\omega_{ij})  & \omega_{ij} > 0
    \end{array}\right. .
\end{equation}
The transitions are either cooling, dephasing, or heating (denoted $J^{(c)}_n$, $J^{(p)}_n$, $J^{(h)}_n$, respectively) depending upon the change in energy.  These operators are:
\begin{equation}
\begin{split}
    J^{(c)}_n(\omega_{ij})  =&\sqrt{ 2 (\bar{n}(\omega_{ij}) +1)  \chi_n(|\omega_{ij}|)}\,  \langle E_i|n \rangle \langle n|E_j\rangle\, |E_i\rangle\langle E_j| \\
    J^{(p)}_n =& \sqrt{2\,k_b T\,\left.\frac{d \chi_n}{d \omega}\right|_{\omega = 0} }\, \sum_i \langle E_i|n \rangle \langle n|E_i\rangle\, |E_i\rangle\langle E_i|\\
    J^{(h)}_n(\omega_{ij})  =& \sqrt{ 2\,\bar{n}(\omega_{ij})\,  \chi_n(|\omega_{ij}|)}\,  \langle E_i|n \rangle \langle n|E_j\rangle\, |E_i\rangle\langle E_j|.
\end{split}
\end{equation}
Here $|n\rangle$ is the excitonic state at site n and $|E_j\rangle$ is an energy eigenstate of $H_{\text{sys}}$. $\bar{n}(\omega)$ is the average thermal occupation of the 1D Bose-Einstein distribution,  $\chi_n(\omega)$ is the one-sided spectral density of the phononic bath coupling for site $n$ at frequency $\omega$. 

For the LHCII trimers and outer minor core complexes we use the same spectral densities as those employed in ref.~\cite{bennett_structure-based_2013}, which are based upon the experimental fits of ref.~\cite{novoderezhkin_intra-_2011}.  However, for the RC and inner core complexes CP-47 and CP-43 we used the spectral densities from ref.~\cite{novoderezhkin_pathways_2005}, in contrast to the more smoothly varying densities of ref. \cite{raszewski_light_2008}.    The distinction is that the former model has a spectral density that more closely matches to that of the other subcomplexes of PSII, whereas the latter smoothly varying density excludes any dephasing interaction $J^{(p)}_n$, since $\left.\frac{d \chi_n}{d \omega}\right|_{\omega = 0} = 0$ for that form.  

\subsection{Radical pair formation}
The ultimate measure for energy transport in PSII is the formation of chemically useful charge separation.  Thus in order to get a measure of the quantum efficiency (QE) of the system, i.e., the probability for creating charge separation given that a photon was absorbed, we add additional incoherent transitions between specific states in the RC and aggregate effective states representing stages of charge separation and radical pair formation.  

This is done with a coarse-grained model by introducing two states, $|\text{RP}_1\rangle$ and $|\text{RP}_2\rangle$, which represent two steps in the charge transfer process~\cite{bennett_structure-based_2013,roden_long-range_2016}. The first, $|\text{RP}_1\rangle$, is able to reversibly exchange an exciton via 3 specific sites in the reaction center and with hopping rates that nominally obeys detailed balance.  $|\text{RP}_1\rangle$ then irreversibly decays into $|\text{RP}_2\rangle$ acting as a final energy sink.  The specific sites that couple to $|\text{RP}_1\rangle$ are the $P_{D_1}$, $P_{D_2}$ special pair, as well as the associated $D_1$ chlorophyll.  All three sites are assumed to have the same transfer rates with $\Gamma_{n \rightarrow RP_1 }^{-1} = 0.64\, \text{ps}$ and $\Gamma_{RP_1 \rightarrow n }^{-1} = 160\, \text{ps}$.   The final irreversible rate from $|\text{RP}_1\rangle$ to $|\text{RP}_2\rangle$ is $\Gamma_{1 \rightarrow 2 }^{-1} = 560\, \text{ps}$. For our chosen temperature $T = 293$ K, and these given rates, detailed balance requires that the energy splitting between $|\text{RP}_1\rangle$ and the reaction center is $\Delta E = 7068\ \text{cm}^{-1}$.  Thus we set the energy of $|RP_1\rangle$ to be $\Delta E$ below the mean of the energy of $P_{D_1}$ and $P_{D_2}$.  Since the irreversible transition to $|RP_2\rangle$ has no up rate, we cannot infer an energy splitting from a detailed balance criterion.  We therefore arbitrarily set $E_{RP_1} - E_{RP_2} = 1.5\ \Delta E$.  

Having made the secular approximation in the exciton-phonon interaction, we will also model the transitions to $|\text{RP}_1\rangle$ as incoherent transitions between energy eigenstates rather than between $|\text{RP}_1\rangle$ and the site basis states (see Eq.~(\ref{eq:phonons_conditional})).  We do so both for mathematical simplicity and consistency, in that if we ignore possible coherence in the phononic transitions by making the secular approximation, we should do the same for a less precise model of charge separation.  With that, our jump operators between the three specific RC sites that couple to $|\text{RP}_1\rangle$, i.e., $n \in \{ P_{D_1},\, P_{D_2},\,  \text{chl D}_1 \}$, are given by
\begin{equation}
\begin{split}\label{eq:RP jumps}
    J_{n}(E_{\text{RP}_1} - E_i ) =& \sqrt{\Gamma_{n \rightarrow RP_1 }}\, \langle n | E_i \rangle\, |\text{RP}_1\rangle \langle E_i | \\
    J_{n}(E_i - E_{\text{RP}_1} ) =& \sqrt{\Gamma_{RP_1 \rightarrow n }}\, \langle E_i | n \rangle\, |E_i\rangle \langle \text{RP}_1 | \\
    J_{RP_1}(E_{\text{RP}_2} - E_{\text{RP}_1} ) =& \sqrt{\Gamma_{1 \rightarrow 2 }}\,  |\text{RP}_2\rangle \langle \text{RP}_1 |.
\end{split}
\end{equation}

Note that the rates implicit in Eq.~(\ref{eq:RP jumps}) do not satisfy detailed balance.  This is because, while we have used the principle of detailed balance to infer $\Delta E$ from the rates of ref. \cite{bennett_structure-based_2013} above,  the transitions in Eq. (\ref{eq:RP jumps}) will only equilibrate to a thermal state if $\langle n | E_i \rangle = \delta_{ni}$, i.e., if the site basis states $|n \rangle$ are also energy eigenstates. This is not possible in the presence of excitonic coupling. In principle, detailed balance could still be maintained if the up and down rates were energy dependent.

\subsection{Non-radiative decay}

The final addition to our model is a non-radiative decay pathway for every excitonic site $n$.
%, in addition to the three radical pair formation pathways discussed above. We do so 
We do this because the non-radiative decay rates can be of the same order of magnitude as spontaneous emission rates, e.g., for PSII the average non-radiative jump rate is
$\gamma_{nr}^{-1} = 2\, \text{ns}$ \cite{bennett_structure-based_2013}. We shall assume all sites have this average value and model the non-radiative decay via the jump operator 
\begin{equation}
    J^{(nr)}_{n} = \sqrt{\gamma_{nr}}\, |f\rangle\langle n |,
\end{equation}
where the state $|f\rangle$ is a state distinct from $|gnd\rangle$ that we introduce to distinguish radiative from non-radiative decay in our quantum master equation evolution.

\subsection{Photon emission simulations}

\begin{figure}
\includegraphics[width=1.0\columnwidth]{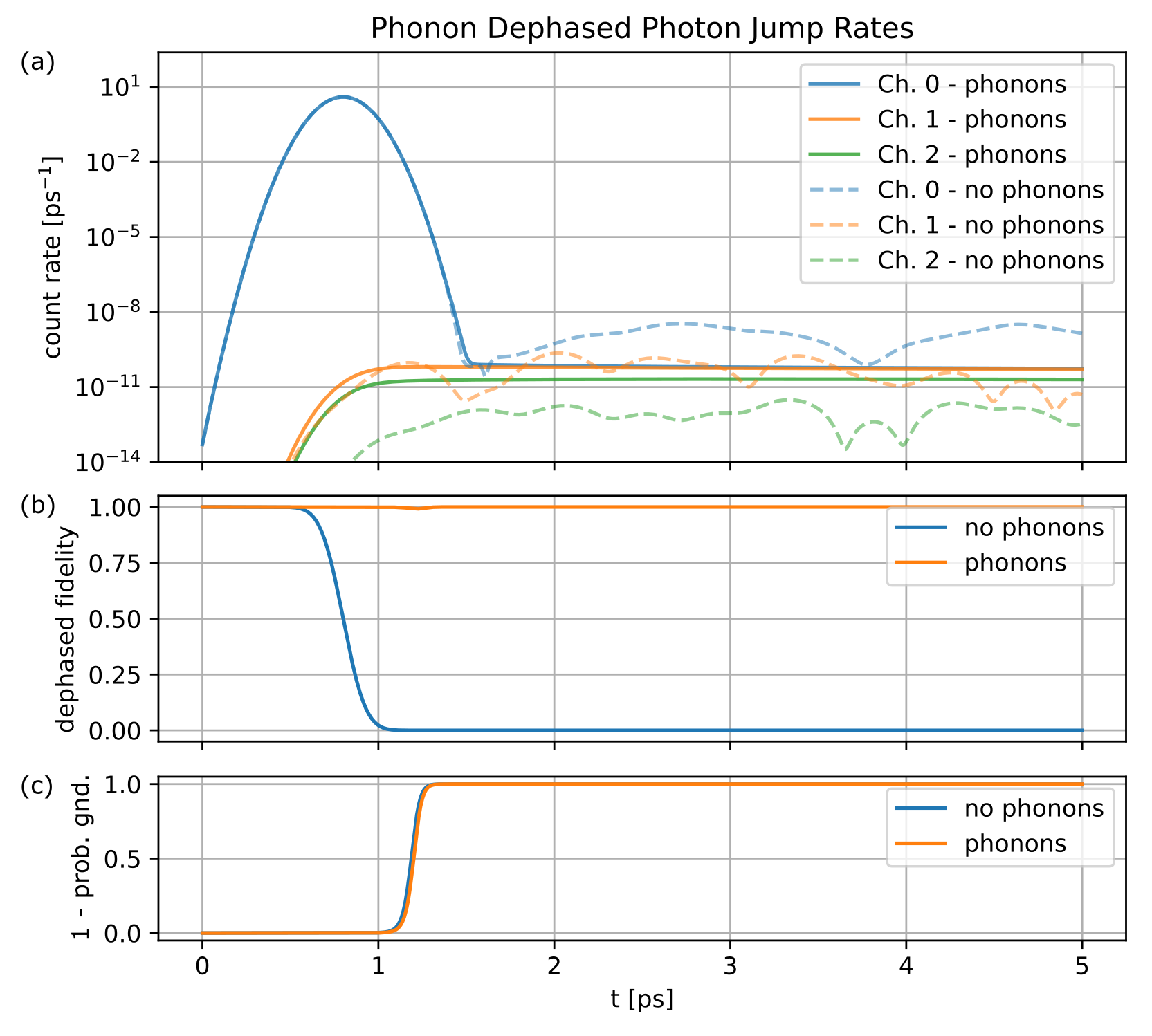}
\caption{ \label{fig:phonons_photon_rates} Comparison of photon emission rates under a strong dephasing phonon model and the no phonon model. (a) Observed single photon count rates in chs. 0-2 for the strong dephasing phononic model (solid lines) and for the no-phonon SSE evolution (dashed lines).
(b) Fidelity of the conditional state with the closest totally dephased state for the strong dephasing model (orange line) and for the no-phonon SSE evolution (blue line).
(c) Probability for PSII to be in any non-ground state, given that no photons were observed, for the strong dephasing phonon model (orange line) and for the no-phonon SSE evolution (blue line).} 
\end{figure}

Under this full model including the RP states and the non-radiative decay, we have computed the probability of emitting a photon into our observed photon channels 0-2.  The expected count rates are contrasted with those from the no phonon model in Fig. \ref{fig:phonons_photon_rates} (a).  We see that the phonon environment has a dramatic damping effect on the oscillations evident in the photon emission rates in Fig. \ref{fig:phonons_photon_rates} (a).  Indeed, after the incident pulse has passed, the phonon model exhibits a nearly constant emission rate for all three emission channels.  In contrast, in the absence of the phonon environment the emission rates vary aperiodically across several orders of magnitude within the 5 ps time frame.

We expect that this large difference is due to the strong dephasing introduced by our particularly simple model of the phonon environment.  This stems from making a secular approximation in the phonon environment, which assumes that every system Bohr frequency $\omega_{nm}$ is on resonance with some particular phonon jump operator and is negligibly off-resonant with all other phonon jump operators. Consequently every peak in the jump rate spectrum of Fig. \ref{fig:spectra} will experience some decoherence, as long as there is a non-zero support in the spectral density.  

To measure the effects of the dephasing in this system we compute the quantum state fidelity, $F(\rho_c, \rho_d) = \operatorname{Tr} \left[\sqrt{\sqrt{\rho_d}\, \rho_c \sqrt{\rho_d} } \right]^2$, between the state conditioned on no photons having been emitted before time $t$, i.e.,  $\rho_c(t)$ and $\rho_d(t)$, the output from a totally dephasing channel, which is given by
\begin{equation}
\rho_d(t) = \sum_{i} \langle E_i |\rho_c(t) | E_i \rangle\,  |E_i \rangle \langle E_i |.
\end{equation}
Fig. \ref{fig:phonons_photon_rates} (b) shows  the quantum state fidelity $F(\rho_c, \rho_d)$ as a function of time for the strongly dephasing phonon model.
For comparison we also show the corresponding  
fidelity for the no phonon model.  We see that when the excitation transitions from being near zero to near one at around $t\sim 1.2$ ps (see Fig.~\ref{fig:cond_0_vs_cond_1}), the fidelity drops only slightly, to a minimum value of $0.991$. This is to be contrasted with the corresponding fidelity for the no-phonon model of Sec.~\ref{sec:simulations} (blue line), where the fidelity falls significantly from unity, due to the fact that in this case the no jump evolution remains a pure state, with significant coherence between energy eigenstates.  This is the case even when the total excited state probability is small.  In fact at the peak arrival time of 0.8 ps, the no-phonon dephasing fidelity already has already fallen to a value of 0.492. 

Fig. \ref{fig:phonons_photon_rates} (c), which plots the probability to be in any non-ground state, shows that despite the difference in dephasing fidelities, the excited state probabilities transition at nearly identical times.
This is because the phononic decoherence sets in only once the system has become excited and acts as the identity channel on the ground state.  Nevertheless, the single photon absorption process remains the same coherent process which feeds the single exciton manifold via the bright state $|B_0\rangle$, with a weighting of this by the incoming photon wave-packet $\xi(t)$ (see Sec.~\ref{sec:emission} and Appendix~\ref{app:paraxial}).  Once that excitonic coherence is created, the phononic decoherence channel will primarily dephase the excitation created at the beginning of the pulse, relative to its later contribution.  However, this dephasing will not impact the total excited state population.  

\subsection{Long time populations in presence of phonons and non-radiative decay}

We now present simulations that compute the average evolution of the system for times $t \gg 1.8$ ps, given that no photonic jump was observed for $t \le 1.8$ ps. Thus for times $t < 1.8 \text{ ps}$ the state will evolve under Eq. (\ref{eq:phonons_conditional}), properly renormalized, while for times $t \ge 1.8 \text{ ps}$ it will follow Eq. (\ref{eq:phonons_unconditional}). This will test our phononic model for long times and give us a numerical estimate for the quantum efficiency, by computing the final population in $|\text{RP}_2\rangle$. We choose 1.8 ps because at this time the single photon pulse is $10\, \sigma$ away from the peak arrival time. Therefore conditioning on observing no photon before 1.8 ps gives us a good criterion that the single photon has been absorbed.

The fidelity analysis above showed that $\rho_c(t)$ remains extremely close to a convex mixture of energy eigenstates throughout its evolution, and in particular by the time we transition to integrating the unconditional master equation of Eq. (\ref{eq:phonons_unconditional}).  It is easy to show that because all of the jump operators map excited states onto energy eigenstates, once the system density matrix is diagonal in the energy basis it will thereafter remain diagonal in this basis.  Therefore, for $t \ge 1.8 \text{ ps}$ we only need to track the diagonal elements of $\rho$ and hence to integrate a purely classical hopping master equation to obtain the populations $P_i = \langle E_i|\rho |E_i \rangle$ at longer times.

\begin{figure}
\includegraphics[width=1.0\columnwidth]{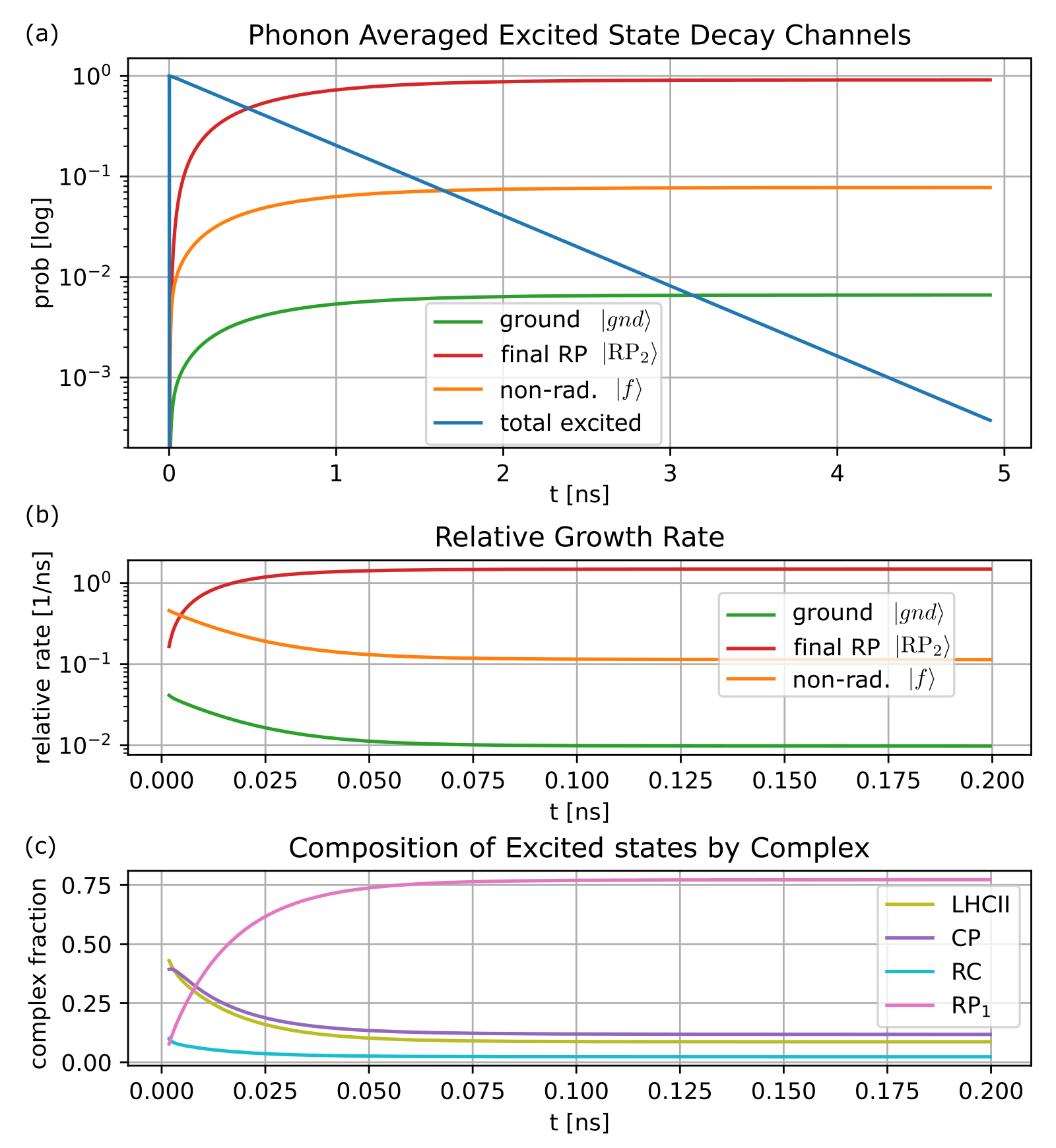}
\caption{ \label{fig:phonons_long_time} Long time population accumulation under the strong dephasing model.  
(a) Average probability for the complex to be in any of the possible terminal states (including the non-stationary excitonic and $|\text{RP}_1\rangle$ states, blue line), given that no photon was emitted in the first 1.8 ps. 
(b) Feeding rates into the terminal states, Eq. (\ref{eq:feeding_rates}), for times $t \ge 1.8 \text{ ps}$.  
(c) Relative fraction of the excited state population distributed between the different PSII sub-complexes and $|\text{RP}_1\rangle$. The relative growth in population of $|\text{RP}_1\rangle$ is consistent with the change in relative rates shown in panel (b). }
\end{figure}

Fig.~\ref{fig:phonons_long_time} (a) shows the results of this simulation for times $0 \le t \le 4.91\text{ ns}$. We track the probability to be in one of the three steady states, $|\text{RP}_2\rangle$, $|f\rangle$, $|gnd\rangle$, corresponding to radical-pair formation, non-radiative decay and single photon emission. In addition, we plot the probability to be in any of the remaining non-stationary excited states, $P_{excited}$, which combines the populations in  $|\text{RP}_1\rangle$ with the population in the single excitation manifold (blue line).  At the final time $t = 4.91\text{ ns}$, the populations are $P_{gnd} = 0.006$, $P_{RP_2} = 0.922$, $P_{f} = 0.072$, and $P_{excited} < 10^{-3}$.
{We note that the final population of 92.2\%  in the radical pair state $|\text{RP}_1\rangle$ constitutes a microscopic estimate of the quantum efficiency, given the single photon nature of the incident pulse and is consistent with experimental measures of the quantum efficiency of PSII~\cite{blankenship_molecular_2014,genty_relationship_1989}.

The dynamics shown in Fig.~\ref{fig:phonons_long_time} can actually be understood by considering an even simpler model than the relatively simple classical hopping between energy eigenstates mentioned above.
A minimal model for population decay is to assume that at the transition time $t = 1.8$ ps, each of the stationary terminal states, i.e., $|\text{RP}_2\rangle$, $|gnd\rangle$,  $|f\rangle$ will be populated by decay from the remaining excited states with a constant and equal rate $\Gamma_i$.  Thus the change in the probability for each eventual stationary state $P_i$ will increase in proportion with rate $\Gamma_i$ to the total probability to be in the excited states, $P_{excited}$.  Because the total probability must be conserved, this implies that $P_{excited}$ must decrease an equal amount in turn and so we have the following simple system of equations,
\begin{equation}
\begin{split} \label{eq:simple_decay}
    \frac{d P_i(t)}{dt} =& \Gamma_i\, P_{excited}(t) \\
    \frac{d P_{excited}(t)}{dt} =& -\sum_i{\Gamma_i} P_{excited}(t).
\end{split}
\end{equation}

To compare our quantum master equation treatment to this minimal model, we compute the time dependent relative rates,
\begin{equation}\label{eq:feeding_rates}
\Gamma_i(t) \equiv \frac{1}{P_{excited}(t)}\, \frac{d P_i(t)}{dt}.  
\end{equation}
If this minimal model were to fit for all time, then each $\Gamma_i$ would be constant in time. Any time dependence in these rates reveals some non-equilibrium dynamics in the decay of $P_{excited}$.  Fig. \ref{fig:phonons_long_time} (b) shows that for times $1.8\text{ ps} \le t \lesssim 100 \text{ ps}$ the rates $\Gamma_i(t)$ are indeed time-dependent and therefore indicate initial transient behavior before full thermalization.  However, for times $t \gtrsim 100 \text{ ps}$ the relative decay rates are essentially flat.  The corresponding asymptotic values are $\Gamma_{gnd}^{-1} = 122 \text{ ns}$, $\Gamma_{RP_2}^{-1} = 0.660 \text{ ns}$, and $\Gamma_{f}^{-1} = 9.45 \text{ ns}$.  From these values we can extract the steady-state population in the radical pair state $|\text{RP}_2\rangle$ from the minimal model of Eq. (\ref{eq:simple_decay}). This analysis also yields a prediction for the QE according to the usual definition based on (macroscopic) rates  $QE = \Gamma_{RP_2}/ (\Gamma_{gnd} + \Gamma_{RP_2} + \Gamma_{f} )$ \cite{blankenship_molecular_2014}. This results in an estimate of 93.0\% for the QE.}  

This estimate is remarkably close to the terminal $|\text{RP}_2\rangle$ population of the full numerical simulations (Fig.\ref{fig:phonons_long_time} (a)).  
Fig. \ref{fig:phonons_long_time} (b) shows that at short times the rate for radical pair production is less than its steady-state value, so it is unsurprising that the minimal model shows a slight over-estimate of the final QE.   
From a microscopic perspective, this transitory behavior can be understood from the fact that the coupling to the incident photon is through the excitonic states, which can decay either radiatively or non-radiatively.  However, in order to irreversibly transfer into $|\text{RP}_2\rangle$, the system must first populate the reversible $|\text{RP}_1\rangle$ state.  In our simple model of radical pair formation, $|\text{RP}_1\rangle$ can only decay to $|\text{RP}_2\rangle$ and cannot decay radiatively or non-radiatively.  Thus as $|\text{RP}_1\rangle$ gains significant population, the probability to be in a state that can decay radiatively or non-radiatively decreases.  Fig. \ref{fig:phonons_long_time} (c) shows the effect of this, plotting how the total excited state population is distributed between the various sub-complexes of PSII and the $|\text{RP}_1 \rangle$ state. 

It is worth noting that in terms of the unit dipole spontaneous emission rate, the radiative rate is $\Gamma_{gnd} = 3.99\, \Gamma_0$.   This is significantly less than the spontaneous emission rates from single chromophores in PSII, which range from  10  to 20  $\Gamma_0$. However, Fig. \ref{fig:phonons_long_time} (c) shows that 75\% of the steady-state excitation is in the non-radiative $|\text{RP}_1\rangle$ state. An estimate of the overall radiative rate at steady-state is then 25\% of a uniform average over every individual emission rate, i.e., $0.25 \frac{1}{N}\sum_{n} \Gamma_n = 3.81\, \Gamma_0$. This is quite close to the thermal steady state value of $3.99\, \Gamma_0$, with the slight bias away from a uniform average attributed to the unequal weighting between the LHCII, CP and RC complexes, which themselves have an unequal distribution of dipoles with different spontaneous emission rates.

\section{Discussion and Conclusions \label{sec:conclusion} }

Here we have derived a quantum trajectory model of PSII interacting with an incident single photon.  In doing so, we have shown how the incident photon couples to a collective dipole state that is distinct from both the Hamiltonian eigenstates and the eigenvectors of the decaying non-Hermitian effective Hamiltonian that generates the smooth evolution between jumps in the quantum trajectories. 
We have also shown what information can be gained by measuring the output photon modes.  In particular, we showed through both numerical simulation and theoretical analysis that the rare events where the input photon is not detected in any output mode results in the excited state probability converging to one as the incident wave-packet passes the system.  We compared this to the case of a coherent state with one mean photon and showed that the near unit excitation probability is not achieved for a coherent state because of its significant overlap with the vacuum state.

Finally, we showed both numerically and analytically how, in the otherwise isolated PSII complex, the spectrum of observed fluorescent photon counts indicates that the initially excited bright state is not a stationary state and that the oscillations in the observed fluorescent count rates can serve as a witness to coherent dynamics in the excited states. 
These studies without exciton-phonon coupling were complemented by studies including excitonic coupling to phonon modes withing a strongly dephasing model of the latter. We have shown here that in a strongly dephasing environment, the fluorescent count rate oscillations are strongly suppressed and essentially absent for the PSII simulations presented here.  In this situation the average state of the system has extremely high fidelity with a density matrix that lacks any coherence between energy eigenstates.  While this is a very interesting and leading result, we note that neither of the two extremes of no phonons and a strongly dephasing phononic environment are truly an accurate representation for the complex phononic environment of photosynthetic excitons.  What role phonons play photosynthetic complexes is still a matter of active research, with some experiments~\cite{arsenault_vibronic_2020} and theoretical studies~\cite{arsenault_vibronic_2021} indicating that some quantum coherence between the phonon environment and the excited system may appear as a result of non-Markovian exciton-phonon couplings.

We expect that a fruitful avenue of research to further investigate this important issue will be to weaken the full secular approximation into partially coherent or `quasi-secular' jump operators, with slowly varying Bohr frequencies retained as a source of coherent dynamics~\cite{tscherbul_partial_2015}. Such partial secular approximations have already been applied to study photosynthetic complexes~\cite{jeske_bloch-redfield_2015,amarnath_multiscale_2016} and we expect they can usefully be applied here.  
In a following paper, we shall combine these techniques with a quantum trajectory picture in which the photon detection is accompanied by phonon detection.  In the same manner that the presence or absence of an exiting photon can provide substantial information about a quantum system of interest, the strength of the dephasing introduced by the Markovian phonon bath studied here suggests that exiting phonons also carry with them a substantial amount of information which can be analyzed theoretically, although accessing such information experimentally would present a significant challenge. 

\subsection{Acknowledgements}
This work was supported by the Photosynthetic Systems program of U.S. Department of Energy, Office of Science, Basic Energy Sciences, under Award No. DE-SC0019728.

\appendix

\section{Paraxial coupling model \label{app:paraxial} }
Here we derive the coupling between the sample and a paraxial field mode and express the coupling effiency factor $\eta$ both in terms of the paraxial beam's cross sectional area, as well as the 98\% collection solid angle. 

Formally, we can decompose the full electric field operator $\mathbf{E}$, in terms of a field operator $\mathbf{E}_\text{para}$ and the electric field for the remaining modes, $\mathbf{E}_{\perp}$:
\begin{equation}
\mathbf{E} = \mathbf{E}^{(+)}_\text{para} + \mathbf{E}_{\perp}^{(+)} + h. c.,
\end{equation}
($h. c.$ stands for the Hermetian conjugate of the preceding terms.) Specifically, the positive frequency component of the freely propagating electric field operator for a TEM$_{00}$ paraxial beam (with carrier wave number $k_0$ propagating in the $+z$ direction) is ~\cite[appendix B]{baragiola_three-dimensional_2014}:
\begin{multline}\label{eq:Eparaxial}
\mathbf{E}^{(+)}_\text{para}(\mathbf{r}, t) = i \sqrt{\frac{\hbar \omega_0}{2 \epsilon_0  A c }} u_{00} (\mathbf{x}_\perp, z) e^{i k_0(z - c t)}\\  \sum_{q = 1,2} \mathbf{e}_q(\mathbf{x}_\perp, z)\, \hat{b}_q(t - z/c),
\end{multline}
where $u_{00}(\mathbf{x}_\perp, z)$ is the unitless scalar mode function for a TEM$_{00}$ beam (square normalized to the transverse area $A$, i.e. $\int d^2 x_\perp\, |u_{00}(\mathbf{x}_\perp, z)|^2 = A\, $ ), and $\mathbf{e}_{q}(\mathbf{x}_\perp, z)$ are the 2 orthogonal polarization states and the field operators $\hat{b}_q$ and $\hat{b}_q^\dag$ obey the delta commutation relation:
\begin{equation}\label{eq:SlowCommutation}
\left[ \hat{b}_q(t - z/c), \hat{b}_{q'}^\dag (t' - z'/c)\right] = \delta_{q q'} \delta (t - t' - (z - z')/c).
\end{equation}
To ensure maximal coupling, we will assume that the beam's carrier frequency is equal to the resonance transition frequency $\omega_0$ and that the chromophore is located at the center of its focus, there by setting $\mathbf{r}_a = \boldsymbol{0}$. 

Taking the paraxial part of the $-\mathbf{E} \cdot \hat{\mathbf{d}}$ interaction, we can see that, in a frame rotating at $\omega_0$ and making the rotating wave approximation, the resonant interaction between a single dipole and the paraxial beam is
\begin{multline}
    - \mathbf{E}_\text{para}(\mathbf{r}_a, t) \cdot \hat{\mathbf{d}}(t) = - i \sum_{q = 1,2}\sqrt{\frac{\hbar \omega_0 d_0^2}{2 \epsilon_0  A c }} u_{00} (\mathbf{x}_{\perp a}, z_a) \\
     e^{i k_0 z_a}\, \mathbf{e}_q(\mathbf{x}_{\perp a}, z_a)\cdot \mathbf{d}\ \hat{\sigma}_{+} \hat{b}_q(t - z_a/c)  + h. c.
\end{multline}
Note that because in the rotating frame $\hat{\mathbf{d}}^{(+)}(t) = e^{+ i \omega_0 t} \sigma_{+}$ this explicit time cancels the time dependence of the carrier, only leaving the time dependence of the slowly-varying field operator $\hat{b}_q(t)$.  This expression can be simplified by introducing the unit Wigner-Weisskopf decay rate $\gamma_0 = \frac{d_0^2 \omega_0^3}{3 \pi \epsilon_0 c^3}$, the approximation that $\mathbf{r}_a = 0$ for all chromophores resulting in the universal geometric factor 
\begin{equation}
\eta \equiv \frac{3 \pi}{2 k_0^2 A} |u_{00}(\boldsymbol{0})|^2. 
\end{equation}
The total interaction between all $N$ sites and the paraxial beam is then
\begin{multline}
    - \mathbf{E}_\text{para}(\mathbf{0}, t) \cdot \sum_{j = 1}^N\hat{\mathbf{d}}_j(t) \approx \\
    - i \hbar\sum_{q = 1,2} \sum^N_{j = 1}  \mathbf{e}_q(\boldsymbol{0}) \cdot \mathbf{d}_j  \,\hat{\sigma}_{+}^{(j)}\ \hat{b}_q(t) + h. c.
\end{multline}
This interaction shows that the operator $\hat{b}_q(t)$ resonantly couples to the system via the collective system operator:
\begin{equation}
L^\dag_{\text{para}\, q} \equiv \sqrt{\gamma_0\, \eta }\, \mathbf{e}_q (\boldsymbol{0}) \cdot \mathbf{D}^\dag, 
\end{equation}
An alternative form for $\eta$  can be found by writing it in terms of the ($98\%$ transmission) collection solid angle $\Delta \Omega$ for a TEM$_{00}$ Gaussian beam.  The beam area is often written in terms of the minimum (1/e) radius of the electric field, call the beam waist $\mathsf{w}_0$, with $A = \pi \mathsf{w}_0 ^2 /2$.  When written in terms of the the beam waist, $\Delta \Omega = 8 \pi /(k_0 \mathsf{w}_0)^2 $, and so we have the equivalent expression $\eta = \tfrac{3 }{8 \pi} \Delta \Omega $ used in the main text.

\section{geometric mode decomposition \label{app:geometry}}
In the main text we consider the special case geometry of two observation spatial modes, with identical collection solid angles and orthogonal propagation directions.  In general multiple beams can be considered so long as they have negligible spatial overlaps, orthogonal polarizations, or both.  Here we consider the case of varying the angle between incident and a collection beam, and the relation between the incident polariziation to the two polarization states available to the collection beam.  For simplicity, we will assume that both the incident paraxial beam and off axis observation modes have the same collection angle $\Delta \Omega$, but that the polarization vectors are generally distinct but not necessary orthogonal.  (If the incident beam propagates in the $+z$ direction and is $+x$ polarized at its focus, an observation beam that propagates along $+y$ could also be polarized along $+x$.)  We denote the polarization vector for incoming beam as $\mathbf{e}_\mathsf{in}$, and the two polarization states of the observation mode as $\mathbf{e}_{\mathsf{obs} 1}$ and $\mathbf{e}_{\mathsf{obs} 2}$.  So while we require $\mathbf{e}_{\mathsf{obs} 1} \cdot \mathbf{e}_{\mathsf{obs} 2} = 0$, neither of these vectors need be orthogonal to $\mathbf{e}_{\mathsf{in}}$, unless $\mathbf{e}_\mathsf{in}$ is also parallel to the observation optical axis.  Thus these three polarization vectors lead to the 3 paraxial coupling operators, 
\begin{equation}
L_{p} = \sqrt{\gamma_0\, \eta }\, \mathbf{e}_p \cdot \mathbf{D}
\end{equation}
where $\mathbf{e}_p \in \{\mathbf{e}_\mathsf{in}, \mathbf{e}_{\mathsf{obs}\,1}, \mathbf{e}_{\mathsf{obs}\, 2} \}$.

The constraint equation for the loss operators, Eq.~(\ref{eq:ch_loss_constraint}) can now be written, such that in general the  set of operators $L_{\perp\, q}$, must satisfy:
\begin{equation}
\sum_{q } L_{\perp\, q}^\dag L_{\perp\, q}  = \gamma_0 \mathbf{D}^\dag \cdot \mathbf{D} -  \sum_{p} L_p^\dag L_p.
\end{equation}
It is illuminating to write this equation in terms of the $3 \times 3$ identity matrix, $\mathbf{I}$, as well as the dyads formed by the unit vectors $\mathbf{e}_p$:
\begin{equation}
\sum_{q } L_{\perp\, q}^\dag L_{\perp\, q}  = \gamma_0 \mathbf{D}^\dag\cdot( \mathbf{I} - \eta \sum_p \mathbf{e}_p \mathbf{e}_p ) \cdot \mathbf{D}.
\end{equation}
In the main text, the center $3 \times 3$ matrix,
\begin{equation}
    \mathbf{M} \equiv \mathbf{I} - \eta \sum_p \mathbf{e}_p \mathbf{e}_p,
\end{equation}
was simply proportional to $\mathbf{I}$, leading to a particularly simple solution.  However, in general $\mathbf{M}$ has nontrivial eigenvalues $\{m_i\}$ and eigenvectors $\{ \mathbf{v}_ i\}$, but once we solve this problem we then have the solution:
\begin{equation}
    L_{\perp\, i} = \sqrt{\gamma_0\, m_i}\, \mathbf{v}_i \cdot \mathbf{D}.
\end{equation}

Solving for these eigenvalues and eigenvectors is easily done.  However, an analytic expression is obtainable, by considering the fact that $\mathbf{e}_{\mathsf{obs}\, 1}$, $\mathbf{e}_{\mathsf{obs}\, 2}$, and the propagation direction of the observation beam $\mathbf{e}_{\mathsf{obs}\, k }$ form an orthonormal basis. This means that $\mathbf{M}$ can be written as,
\begin{equation}
\mathbf{M}  = (1 - \eta)\, \mathbf{I} - \eta \, (\mathbf{e}_\mathsf{in} \mathbf{e}_\mathsf{in} - \mathbf{e}_{\mathsf{obs}\, k } \mathbf{e}_{\mathsf{obs}\, k }).
\end{equation}
Thus when $\mathbf{e}_\mathsf{in}  \cdot \mathbf{e}_{\mathsf{obs}\, k } = \pm 1$, $\mathbf{M}$ is proportional to the identity.  Also, note that any vector that is orthogonal to both $\mathbf{e}_\mathsf{in}$ and $\mathbf{e}_{\mathsf{obs}\, k }$ is an eigenvector of $\mathbf{M}$, with eigenvalue $1 - \eta$.  

We define $\theta$ to be the angle between $\mathbf{e}_\mathsf{in}$ and $\mathbf{e}_{\mathsf{obs}\, 1}$ - $\mathbf{e}_{\mathsf{obs}\, 2}$ plane, which is also equal to $\theta = \sin^{-1}(\mathbf{e}_\mathsf{in}  \cdot \mathbf{e}_{\mathsf{obs}\, k } )$.  Then the three eigenvalues of $\mathbf{M}$ turn out to be $m_q = 1 - \eta \,( 1 +  q \, \cos(\theta) )$ where $q = -1, 0, 1$.  Whenever $\cos (\theta) \ne 0$, the $q = 0$ eigenvector is equal to,
\begin{equation}
    \mathbf{v}_0 = (\mathbf{e}_{\mathsf{in}} \times \mathbf{e}_{\mathsf{obs}\, k} )/\cos (\theta).
\end{equation}
The eigenvector $\mathbf{v}_{+1}$ defines the dipole orientation that maximally couples to all of the observed modes, thereby minimizing the coupling to the unobserved modes. It turns out that this is the vector that lies half way between the input polarization $\mathbf{e}_{\mathsf{in}}$, and its projection into the $\mathbf{e}_{\mathsf{obs} 1}$ - $\mathbf{e}_{\mathsf{obs} 2} $ plane.  It then follows that a dipole oriented along $\mathbf{v}_{-1}$ is minimally coupled to the observed modes, and is equal to $ \mathbf{v}_{-1} = \mathbf{v}_{0} \times \mathbf{v}_{+1}$.  The degenerate case of $\theta = \pi/2$ is the geometry we consider in the main text and $\boldsymbol{\mathsf{M}} \propto \mathbf{I}$.  

\bibliography{PSII}

\begin{thebibliography}{52}
\expandafter\ifx\csname natexlab\endcsname\relax\def\natexlab#1{#1}\fi
\expandafter\ifx\csname bibnamefont\endcsname\relax
  \def\bibnamefont#1{#1}\fi
\expandafter\ifx\csname bibfnamefont\endcsname\relax
  \def\bibfnamefont#1{#1}\fi
\expandafter\ifx\csname citenamefont\endcsname\relax
  \def\citenamefont#1{#1}\fi
\expandafter\ifx\csname url\endcsname\relax
  \def\url#1{\texttt{#1}}\fi
\expandafter\ifx\csname urlprefix\endcsname\relax\def\urlprefix{URL }\fi
\providecommand{\bibinfo}[2]{#2}
\providecommand{\eprint}[2][]{\url{#2}}

\bibitem[{\citenamefont{Blankenship}(2014)}]{blankenship_molecular_2014}
\bibinfo{author}{\bibfnamefont{R.~E.} \bibnamefont{Blankenship}},
  \emph{\bibinfo{title}{Molecular {{Mechanisms}} of {{Photosynthesis}}, 2nd
  {{Edition}}}} (\bibinfo{publisher}{{Wiley-Blackwell}},
  \bibinfo{address}{{Chichester, West Sussex}}, \bibinfo{year}{2014}),
  \bibinfo{edition}{2nd} ed.

\bibitem[{\citenamefont{Genty et~al.}(1989)\citenamefont{Genty, Briantais, and
  Baker}}]{genty_relationship_1989}
\bibinfo{author}{\bibfnamefont{B.}~\bibnamefont{Genty}},
  \bibinfo{author}{\bibfnamefont{J.-M.} \bibnamefont{Briantais}},
  \bibnamefont{and} \bibinfo{author}{\bibfnamefont{N.~R.} \bibnamefont{Baker}},
  \bibinfo{journal}{Biochimica et Biophysica Acta (BBA) - General Subjects}
  \textbf{\bibinfo{volume}{990}}, \bibinfo{pages}{87} (\bibinfo{year}{1989}).

\bibitem[{\citenamefont{Brixner et~al.}(2005)\citenamefont{Brixner, Stenger,
  Vaswani, Cho, Blankenship, and Fleming}}]{brixner_two-dimensional_2005}
\bibinfo{author}{\bibfnamefont{T.}~\bibnamefont{Brixner}},
  \bibinfo{author}{\bibfnamefont{J.}~\bibnamefont{Stenger}},
  \bibinfo{author}{\bibfnamefont{H.~M.} \bibnamefont{Vaswani}},
  \bibinfo{author}{\bibfnamefont{M.}~\bibnamefont{Cho}},
  \bibinfo{author}{\bibfnamefont{R.~E.} \bibnamefont{Blankenship}},
  \bibnamefont{and} \bibinfo{author}{\bibfnamefont{G.~R.}
  \bibnamefont{Fleming}}, \bibinfo{journal}{Nature}
  \textbf{\bibinfo{volume}{434}}, \bibinfo{pages}{625} (\bibinfo{year}{2005}).

\bibitem[{\citenamefont{Engel et~al.}(2007)\citenamefont{Engel, Calhoun, Read,
  Ahn, Man{\v c}al, Cheng, Blankenship, and Fleming}}]{engel_evidence_2007}
\bibinfo{author}{\bibfnamefont{G.~S.} \bibnamefont{Engel}},
  \bibinfo{author}{\bibfnamefont{T.~R.} \bibnamefont{Calhoun}},
  \bibinfo{author}{\bibfnamefont{E.~L.} \bibnamefont{Read}},
  \bibinfo{author}{\bibfnamefont{T.-K.} \bibnamefont{Ahn}},
  \bibinfo{author}{\bibfnamefont{T.}~\bibnamefont{Man{\v c}al}},
  \bibinfo{author}{\bibfnamefont{Y.-C.} \bibnamefont{Cheng}},
  \bibinfo{author}{\bibfnamefont{R.~E.} \bibnamefont{Blankenship}},
  \bibnamefont{and} \bibinfo{author}{\bibfnamefont{G.~R.}
  \bibnamefont{Fleming}}, \bibinfo{journal}{Nature}
  \textbf{\bibinfo{volume}{446}}, \bibinfo{pages}{782} (\bibinfo{year}{2007}).

\bibitem[{\citenamefont{Wang et~al.}(2019)\citenamefont{Wang, Allodi, and
  Engel}}]{wang_quantum_2019}
\bibinfo{author}{\bibfnamefont{L.}~\bibnamefont{Wang}},
  \bibinfo{author}{\bibfnamefont{M.~A.} \bibnamefont{Allodi}},
  \bibnamefont{and} \bibinfo{author}{\bibfnamefont{G.~S.} \bibnamefont{Engel}},
  \bibinfo{journal}{Nat Rev Chem} \textbf{\bibinfo{volume}{3}},
  \bibinfo{pages}{477} (\bibinfo{year}{2019}).

\bibitem[{\citenamefont{Cao et~al.}(2020)\citenamefont{Cao, Cogdell, Coker,
  Duan, Hauer, Kleinekath{\"o}fer, Jansen, Man{\v c}al, Miller, Ogilvie
  et~al.}}]{cao_quantum_2020}
\bibinfo{author}{\bibfnamefont{J.}~\bibnamefont{Cao}},
  \bibinfo{author}{\bibfnamefont{R.~J.} \bibnamefont{Cogdell}},
  \bibinfo{author}{\bibfnamefont{D.~F.} \bibnamefont{Coker}},
  \bibinfo{author}{\bibfnamefont{H.-G.} \bibnamefont{Duan}},
  \bibinfo{author}{\bibfnamefont{J.}~\bibnamefont{Hauer}},
  \bibinfo{author}{\bibfnamefont{U.}~\bibnamefont{Kleinekath{\"o}fer}},
  \bibinfo{author}{\bibfnamefont{T.~L.~C.} \bibnamefont{Jansen}},
  \bibinfo{author}{\bibfnamefont{T.}~\bibnamefont{Man{\v c}al}},
  \bibinfo{author}{\bibfnamefont{R.~J.~D.} \bibnamefont{Miller}},
  \bibinfo{author}{\bibfnamefont{J.~P.} \bibnamefont{Ogilvie}},
  \bibnamefont{et~al.}, \bibinfo{journal}{Sci Adv}
  \textbf{\bibinfo{volume}{6}}, \bibinfo{pages}{eaaz4888}
  (\bibinfo{year}{2020}).

\bibitem[{\citenamefont{Chan et~al.}(2018)\citenamefont{Chan, Gamel, Fleming,
  and Whaley}}]{chan_single-photon_2018}
\bibinfo{author}{\bibfnamefont{H.~C.~H.} \bibnamefont{Chan}},
  \bibinfo{author}{\bibfnamefont{O.~E.} \bibnamefont{Gamel}},
  \bibinfo{author}{\bibfnamefont{G.~R.} \bibnamefont{Fleming}},
  \bibnamefont{and} \bibinfo{author}{\bibfnamefont{K.~B.}
  \bibnamefont{Whaley}}, \bibinfo{journal}{J. Phys. B: At. Mol. Opt. Phys.}
  \textbf{\bibinfo{volume}{51}}, \bibinfo{pages}{054002}
  (\bibinfo{year}{2018}).

\bibitem[{\citenamefont{Kaneda et~al.}(2016)\citenamefont{Kaneda,
  {Garay-Palmett}, U'Ren, and Kwiat}}]{kaneda_heralded_2016}
\bibinfo{author}{\bibfnamefont{F.}~\bibnamefont{Kaneda}},
  \bibinfo{author}{\bibfnamefont{K.}~\bibnamefont{{Garay-Palmett}}},
  \bibinfo{author}{\bibfnamefont{A.~B.} \bibnamefont{U'Ren}}, \bibnamefont{and}
  \bibinfo{author}{\bibfnamefont{P.~G.} \bibnamefont{Kwiat}},
  \bibinfo{journal}{Opt. Express, OE} \textbf{\bibinfo{volume}{24}},
  \bibinfo{pages}{10733} (\bibinfo{year}{2016}).

\bibitem[{\citenamefont{Phan et~al.}(2014)\citenamefont{Phan, Cheng, Bessarab,
  and Krivitsky}}]{phan_interaction_2014}
\bibinfo{author}{\bibfnamefont{N.~M.} \bibnamefont{Phan}},
  \bibinfo{author}{\bibfnamefont{M.~F.} \bibnamefont{Cheng}},
  \bibinfo{author}{\bibfnamefont{D.~A.} \bibnamefont{Bessarab}},
  \bibnamefont{and} \bibinfo{author}{\bibfnamefont{L.~A.}
  \bibnamefont{Krivitsky}}, \bibinfo{journal}{Phys. Rev. Lett.}
  \textbf{\bibinfo{volume}{112}}, \bibinfo{pages}{213601}
  (\bibinfo{year}{2014}).

\bibitem[{\citenamefont{Bakulin et~al.}(2016)\citenamefont{Bakulin, Silva, and
  Vella}}]{bakulin_ultrafast_2016}
\bibinfo{author}{\bibfnamefont{A.~A.} \bibnamefont{Bakulin}},
  \bibinfo{author}{\bibfnamefont{C.}~\bibnamefont{Silva}}, \bibnamefont{and}
  \bibinfo{author}{\bibfnamefont{E.}~\bibnamefont{Vella}}, \bibinfo{journal}{J.
  Phys. Chem. Lett.} \textbf{\bibinfo{volume}{7}}, \bibinfo{pages}{250}
  (\bibinfo{year}{2016}).

\bibitem[{\citenamefont{Ishizaki et~al.}(2010)\citenamefont{Ishizaki, Calhoun,
  {Schlau-Cohen}, and Fleming}}]{ishizaki_quantum_2010}
\bibinfo{author}{\bibfnamefont{A.}~\bibnamefont{Ishizaki}},
  \bibinfo{author}{\bibfnamefont{T.~R.} \bibnamefont{Calhoun}},
  \bibinfo{author}{\bibfnamefont{G.~S.} \bibnamefont{{Schlau-Cohen}}},
  \bibnamefont{and} \bibinfo{author}{\bibfnamefont{G.~R.}
  \bibnamefont{Fleming}}, \bibinfo{journal}{Phys. Chem. Chem. Phys.}
  \textbf{\bibinfo{volume}{12}}, \bibinfo{pages}{7319} (\bibinfo{year}{2010}).

\bibitem[{\citenamefont{Cook et~al.}()\citenamefont{Cook, Ko, and
  Whaley}}]{cook_phonons_2022}
\bibinfo{author}{\bibfnamefont{R.~L.} \bibnamefont{Cook}},
  \bibinfo{author}{\bibfnamefont{L.}~\bibnamefont{Ko}}, \bibnamefont{and}
  \bibinfo{author}{\bibfnamefont{K.~B.} \bibnamefont{Whaley}},
  \bibinfo{note}{in preparation}.

\bibitem[{\citenamefont{Carmichael and Kim}(2000)}]{carmichael_quantum_2000}
\bibinfo{author}{\bibfnamefont{H.~J.} \bibnamefont{Carmichael}}
  \bibnamefont{and} \bibinfo{author}{\bibfnamefont{K.}~\bibnamefont{Kim}},
  \bibinfo{journal}{Optics Communications} \textbf{\bibinfo{volume}{179}},
  \bibinfo{pages}{417} (\bibinfo{year}{2000}).

\bibitem[{\citenamefont{Plenio and Knight}(1998)}]{plenio_quantum-jump_1998}
\bibinfo{author}{\bibfnamefont{M.~B.} \bibnamefont{Plenio}} \bibnamefont{and}
  \bibinfo{author}{\bibfnamefont{P.~L.} \bibnamefont{Knight}},
  \bibinfo{journal}{Rev. Mod. Phys.} \textbf{\bibinfo{volume}{70}},
  \bibinfo{pages}{101} (\bibinfo{year}{1998}).

\bibitem[{\citenamefont{Loudon}(2000)}]{loudon_quantum_2000}
\bibinfo{author}{\bibfnamefont{R.}~\bibnamefont{Loudon}},
  \emph{\bibinfo{title}{The Quantum Theory of Light}}
  (\bibinfo{publisher}{{Oxford University Press}}, \bibinfo{year}{2000}).

\bibitem[{\citenamefont{Ruskov et~al.}(2007)\citenamefont{Ruskov, Mizel, and
  Korotkov}}]{ruskov_crossover_2007}
\bibinfo{author}{\bibfnamefont{R.}~\bibnamefont{Ruskov}},
  \bibinfo{author}{\bibfnamefont{A.}~\bibnamefont{Mizel}}, \bibnamefont{and}
  \bibinfo{author}{\bibfnamefont{A.~N.} \bibnamefont{Korotkov}},
  \bibinfo{journal}{Phys. Rev. B} \textbf{\bibinfo{volume}{75}},
  \bibinfo{pages}{220501(R)} (\bibinfo{year}{2007}).

\bibitem[{\citenamefont{Minev et~al.}(2019)\citenamefont{Minev, Mundhada,
  Shankar, Reinhold, {Guti{\'e}rrez-J{\'a}uregui}, Schoelkopf, Mirrahimi,
  Carmichael, and Devoret}}]{minev_catch_2019}
\bibinfo{author}{\bibfnamefont{Z.~K.} \bibnamefont{Minev}},
  \bibinfo{author}{\bibfnamefont{S.~O.} \bibnamefont{Mundhada}},
  \bibinfo{author}{\bibfnamefont{S.}~\bibnamefont{Shankar}},
  \bibinfo{author}{\bibfnamefont{P.}~\bibnamefont{Reinhold}},
  \bibinfo{author}{\bibfnamefont{R.}~\bibnamefont{{Guti{\'e}rrez-J{\'a}uregui}}},
  \bibinfo{author}{\bibfnamefont{R.~J.} \bibnamefont{Schoelkopf}},
  \bibinfo{author}{\bibfnamefont{M.}~\bibnamefont{Mirrahimi}},
  \bibinfo{author}{\bibfnamefont{H.~J.} \bibnamefont{Carmichael}},
  \bibnamefont{and} \bibinfo{author}{\bibfnamefont{M.~H.}
  \bibnamefont{Devoret}}, \bibinfo{journal}{Nature}
  \textbf{\bibinfo{volume}{570}}, \bibinfo{pages}{200} (\bibinfo{year}{2019}).

\bibitem[{\citenamefont{Dicke}(1954)}]{dicke_coherence_1954}
\bibinfo{author}{\bibfnamefont{R.~H.} \bibnamefont{Dicke}},
  \bibinfo{journal}{Phys. Rev.} \textbf{\bibinfo{volume}{93}},
  \bibinfo{pages}{99} (\bibinfo{year}{1954}).

\bibitem[{\citenamefont{Hammerer et~al.}(2010)\citenamefont{Hammerer,
  S{\o}rensen, and Polzik}}]{hammerer_quantum_2010}
\bibinfo{author}{\bibfnamefont{K.}~\bibnamefont{Hammerer}},
  \bibinfo{author}{\bibfnamefont{A.~S.} \bibnamefont{S{\o}rensen}},
  \bibnamefont{and} \bibinfo{author}{\bibfnamefont{E.~S.}
  \bibnamefont{Polzik}}, \bibinfo{journal}{Rev. Mod. Phys.}
  \textbf{\bibinfo{volume}{82}}, \bibinfo{pages}{1041} (\bibinfo{year}{2010}).

\bibitem[{\citenamefont{Duan et~al.}(2001)\citenamefont{Duan, Lukin, Cirac, and
  Zoller}}]{duan_long-distance_2001}
\bibinfo{author}{\bibfnamefont{L.-M.} \bibnamefont{Duan}},
  \bibinfo{author}{\bibfnamefont{M.~D.} \bibnamefont{Lukin}},
  \bibinfo{author}{\bibfnamefont{J.~I.} \bibnamefont{Cirac}}, \bibnamefont{and}
  \bibinfo{author}{\bibfnamefont{P.}~\bibnamefont{Zoller}},
  \bibinfo{journal}{Nature} \textbf{\bibinfo{volume}{414}},
  \bibinfo{pages}{413} (\bibinfo{year}{2001}).

\bibitem[{\citenamefont{Gorshkov et~al.}(2007)\citenamefont{Gorshkov,
  Andr{\'e}, Fleischhauer, S{\o}rensen, and Lukin}}]{gorshkov_universal_2007}
\bibinfo{author}{\bibfnamefont{A.~V.} \bibnamefont{Gorshkov}},
  \bibinfo{author}{\bibfnamefont{A.}~\bibnamefont{Andr{\'e}}},
  \bibinfo{author}{\bibfnamefont{M.}~\bibnamefont{Fleischhauer}},
  \bibinfo{author}{\bibfnamefont{A.~S.} \bibnamefont{S{\o}rensen}},
  \bibnamefont{and} \bibinfo{author}{\bibfnamefont{M.~D.} \bibnamefont{Lukin}},
  \bibinfo{journal}{Phys. Rev. Lett.} \textbf{\bibinfo{volume}{98}},
  \bibinfo{pages}{123601} (\bibinfo{year}{2007}).

\bibitem[{\citenamefont{Novoderezhkin et~al.}(1999)\citenamefont{Novoderezhkin,
  Monshouwer, and {van Grondelle}}}]{novoderezhkin_exciton_1999}
\bibinfo{author}{\bibfnamefont{V.}~\bibnamefont{Novoderezhkin}},
  \bibinfo{author}{\bibfnamefont{R.}~\bibnamefont{Monshouwer}},
  \bibnamefont{and} \bibinfo{author}{\bibfnamefont{R.}~\bibnamefont{{van
  Grondelle}}}, \bibinfo{journal}{J. Phys. Chem. B}
  \textbf{\bibinfo{volume}{103}}, \bibinfo{pages}{10540}
  (\bibinfo{year}{1999}).

\bibitem[{\citenamefont{Savikhin et~al.}(1998)\citenamefont{Savikhin, Buck,
  Struve, Blankenship, Taisova, Novoderezhkin, and
  Fetisova}}]{savikhin_excitation_1998}
\bibinfo{author}{\bibfnamefont{S.}~\bibnamefont{Savikhin}},
  \bibinfo{author}{\bibfnamefont{D.~R.} \bibnamefont{Buck}},
  \bibinfo{author}{\bibfnamefont{W.~S.} \bibnamefont{Struve}},
  \bibinfo{author}{\bibfnamefont{R.~E.} \bibnamefont{Blankenship}},
  \bibinfo{author}{\bibfnamefont{A.~S.} \bibnamefont{Taisova}},
  \bibinfo{author}{\bibfnamefont{V.~I.} \bibnamefont{Novoderezhkin}},
  \bibnamefont{and} \bibinfo{author}{\bibfnamefont{Z.~G.}
  \bibnamefont{Fetisova}}, \bibinfo{journal}{FEBS Letters}
  \textbf{\bibinfo{volume}{430}}, \bibinfo{pages}{323} (\bibinfo{year}{1998}).

\bibitem[{\citenamefont{Malina et~al.}(2021)\citenamefont{Malina, Koehorst,
  B{\'i}na, P{\v s}en{\v c}{\'i}k, and {van
  Amerongen}}}]{malina_superradiance_2021}
\bibinfo{author}{\bibfnamefont{T.}~\bibnamefont{Malina}},
  \bibinfo{author}{\bibfnamefont{R.}~\bibnamefont{Koehorst}},
  \bibinfo{author}{\bibfnamefont{D.}~\bibnamefont{B{\'i}na}},
  \bibinfo{author}{\bibfnamefont{J.}~\bibnamefont{P{\v s}en{\v c}{\'i}k}},
  \bibnamefont{and} \bibinfo{author}{\bibfnamefont{H.}~\bibnamefont{{van
  Amerongen}}}, \bibinfo{journal}{Sci Rep} \textbf{\bibinfo{volume}{11}},
  \bibinfo{pages}{8354} (\bibinfo{year}{2021}).

\bibitem[{\citenamefont{Yakovlev et~al.}(2002)\citenamefont{Yakovlev,
  Novoderezhkin, Taisova, and Fetisova}}]{yakovlev_exciton_2002}
\bibinfo{author}{\bibfnamefont{A.}~\bibnamefont{Yakovlev}},
  \bibinfo{author}{\bibfnamefont{V.}~\bibnamefont{Novoderezhkin}},
  \bibinfo{author}{\bibfnamefont{A.}~\bibnamefont{Taisova}}, \bibnamefont{and}
  \bibinfo{author}{\bibfnamefont{Z.}~\bibnamefont{Fetisova}},
  \bibinfo{journal}{Photosynthesis Research} \textbf{\bibinfo{volume}{71}},
  \bibinfo{pages}{19} (\bibinfo{year}{2002}).

\bibitem[{\citenamefont{Monshouwer et~al.}(1997)\citenamefont{Monshouwer,
  Abrahamsson, {van Mourik}, and {van
  Grondelle}}}]{monshouwer_superradiance_1997}
\bibinfo{author}{\bibfnamefont{R.}~\bibnamefont{Monshouwer}},
  \bibinfo{author}{\bibfnamefont{M.}~\bibnamefont{Abrahamsson}},
  \bibinfo{author}{\bibfnamefont{F.}~\bibnamefont{{van Mourik}}},
  \bibnamefont{and} \bibinfo{author}{\bibfnamefont{R.}~\bibnamefont{{van
  Grondelle}}}, \bibinfo{journal}{J. Phys. Chem. B}
  \textbf{\bibinfo{volume}{101}}, \bibinfo{pages}{7241} (\bibinfo{year}{1997}).

\bibitem[{\citenamefont{K{\"u}hn and
  Sundstr{\"o}m}(1997)}]{kuhn_pumpprobe_1997}
\bibinfo{author}{\bibfnamefont{O.}~\bibnamefont{K{\"u}hn}} \bibnamefont{and}
  \bibinfo{author}{\bibfnamefont{V.}~\bibnamefont{Sundstr{\"o}m}},
  \bibinfo{journal}{J. Chem. Phys.} \textbf{\bibinfo{volume}{107}},
  \bibinfo{pages}{4154} (\bibinfo{year}{1997}).

\bibitem[{\citenamefont{Mohseni et~al.}(2008)\citenamefont{Mohseni, Rebentrost,
  Lloyd, and {Aspuru-Guzik}}}]{mohseni_environment-assisted_2008}
\bibinfo{author}{\bibfnamefont{M.}~\bibnamefont{Mohseni}},
  \bibinfo{author}{\bibfnamefont{P.}~\bibnamefont{Rebentrost}},
  \bibinfo{author}{\bibfnamefont{S.}~\bibnamefont{Lloyd}}, \bibnamefont{and}
  \bibinfo{author}{\bibfnamefont{A.}~\bibnamefont{{Aspuru-Guzik}}},
  \bibinfo{journal}{J. Chem. Phys.} \textbf{\bibinfo{volume}{129}},
  \bibinfo{pages}{174106} (\bibinfo{year}{2008}).

\bibitem[{\citenamefont{Plenio and
  Huelga}(2008)}]{plenio_dephasing-assisted_2008}
\bibinfo{author}{\bibfnamefont{M.~B.} \bibnamefont{Plenio}} \bibnamefont{and}
  \bibinfo{author}{\bibfnamefont{S.~F.} \bibnamefont{Huelga}},
  \bibinfo{journal}{New J. Phys.} \textbf{\bibinfo{volume}{10}},
  \bibinfo{pages}{113019} (\bibinfo{year}{2008}).

\bibitem[{\citenamefont{Amarnath et~al.}(2016)\citenamefont{Amarnath, Bennett,
  Schneider, and Fleming}}]{amarnath_multiscale_2016}
\bibinfo{author}{\bibfnamefont{K.}~\bibnamefont{Amarnath}},
  \bibinfo{author}{\bibfnamefont{D.~I.~G.} \bibnamefont{Bennett}},
  \bibinfo{author}{\bibfnamefont{A.~R.} \bibnamefont{Schneider}},
  \bibnamefont{and} \bibinfo{author}{\bibfnamefont{G.~R.}
  \bibnamefont{Fleming}}, \bibinfo{journal}{PNAS}
  \textbf{\bibinfo{volume}{113}}, \bibinfo{pages}{1156} (\bibinfo{year}{2016}).

\bibitem[{\citenamefont{Bennett et~al.}(2013)\citenamefont{Bennett, Amarnath,
  and Fleming}}]{bennett_structure-based_2013}
\bibinfo{author}{\bibfnamefont{D.~I.~G.} \bibnamefont{Bennett}},
  \bibinfo{author}{\bibfnamefont{K.}~\bibnamefont{Amarnath}}, \bibnamefont{and}
  \bibinfo{author}{\bibfnamefont{G.~R.} \bibnamefont{Fleming}},
  \bibinfo{journal}{J. Am. Chem. Soc.} \textbf{\bibinfo{volume}{135}},
  \bibinfo{pages}{9164} (\bibinfo{year}{2013}).

\bibitem[{\citenamefont{Gross and Haroche}(1982)}]{gross_superradiance_1982}
\bibinfo{author}{\bibfnamefont{M.}~\bibnamefont{Gross}} \bibnamefont{and}
  \bibinfo{author}{\bibfnamefont{S.}~\bibnamefont{Haroche}},
  \bibinfo{journal}{Physics Reports} \textbf{\bibinfo{volume}{93}},
  \bibinfo{pages}{301} (\bibinfo{year}{1982}).

\bibitem[{\citenamefont{Lehmberg}(1970)}]{lehmberg_radiation_1970}
\bibinfo{author}{\bibfnamefont{R.~H.} \bibnamefont{Lehmberg}},
  \bibinfo{journal}{Phys. Rev. A} \textbf{\bibinfo{volume}{2}},
  \bibinfo{pages}{883} (\bibinfo{year}{1970}).

\bibitem[{\citenamefont{Glauber and Lewenstein}(1991)}]{glauber_quantum_1991}
\bibinfo{author}{\bibfnamefont{R.~J.} \bibnamefont{Glauber}} \bibnamefont{and}
  \bibinfo{author}{\bibfnamefont{M.}~\bibnamefont{Lewenstein}},
  \bibinfo{journal}{Phys. Rev. A} \textbf{\bibinfo{volume}{43}},
  \bibinfo{pages}{467} (\bibinfo{year}{1991}).

\bibitem[{\citenamefont{Umena et~al.}(2011)\citenamefont{Umena, Kawakami, Shen,
  and Kamiya}}]{umena_crystal_2011}
\bibinfo{author}{\bibfnamefont{Y.}~\bibnamefont{Umena}},
  \bibinfo{author}{\bibfnamefont{K.}~\bibnamefont{Kawakami}},
  \bibinfo{author}{\bibfnamefont{J.-R.} \bibnamefont{Shen}}, \bibnamefont{and}
  \bibinfo{author}{\bibfnamefont{N.}~\bibnamefont{Kamiya}},
  \bibinfo{journal}{Nature} \textbf{\bibinfo{volume}{473}}, \bibinfo{pages}{55}
  (\bibinfo{year}{2011}).

\bibitem[{\citenamefont{Liu et~al.}(2004)\citenamefont{Liu, Yan, Wang, Kuang,
  Zhang, Gui, An, and Chang}}]{liu_crystal_2004}
\bibinfo{author}{\bibfnamefont{Z.}~\bibnamefont{Liu}},
  \bibinfo{author}{\bibfnamefont{H.}~\bibnamefont{Yan}},
  \bibinfo{author}{\bibfnamefont{K.}~\bibnamefont{Wang}},
  \bibinfo{author}{\bibfnamefont{T.}~\bibnamefont{Kuang}},
  \bibinfo{author}{\bibfnamefont{J.}~\bibnamefont{Zhang}},
  \bibinfo{author}{\bibfnamefont{L.}~\bibnamefont{Gui}},
  \bibinfo{author}{\bibfnamefont{X.}~\bibnamefont{An}}, \bibnamefont{and}
  \bibinfo{author}{\bibfnamefont{W.}~\bibnamefont{Chang}},
  \bibinfo{journal}{Nature} \textbf{\bibinfo{volume}{428}},
  \bibinfo{pages}{287} (\bibinfo{year}{2004}).

\bibitem[{\citenamefont{Wiseman and Milburn}(2010)}]{wiseman_quantum_2010}
\bibinfo{author}{\bibfnamefont{H.~M.} \bibnamefont{Wiseman}} \bibnamefont{and}
  \bibinfo{author}{\bibfnamefont{G.~J.} \bibnamefont{Milburn}},
  \emph{\bibinfo{title}{Quantum {{Measurement}} and {{Control}}}}
  (\bibinfo{publisher}{{Cambridge University Press}}, \bibinfo{year}{2010}).

\bibitem[{\citenamefont{U'Ren et~al.}(2005)\citenamefont{U'Ren, Silberhorn,
  Banaszek, Walmsley, Erdmann, Grice, and Raymer}}]{uren_generation_2005}
\bibinfo{author}{\bibfnamefont{A.~B.} \bibnamefont{U'Ren}},
  \bibinfo{author}{\bibfnamefont{C.}~\bibnamefont{Silberhorn}},
  \bibinfo{author}{\bibfnamefont{K.}~\bibnamefont{Banaszek}},
  \bibinfo{author}{\bibfnamefont{I.~A.} \bibnamefont{Walmsley}},
  \bibinfo{author}{\bibfnamefont{R.}~\bibnamefont{Erdmann}},
  \bibinfo{author}{\bibfnamefont{W.~P.} \bibnamefont{Grice}}, \bibnamefont{and}
  \bibinfo{author}{\bibfnamefont{M.~G.} \bibnamefont{Raymer}},
  \bibinfo{journal}{Laser Phys.} \textbf{\bibinfo{volume}{15}},
  \bibinfo{pages}{16} (\bibinfo{year}{2005}).

\bibitem[{\citenamefont{Gough et~al.}(2011)\citenamefont{Gough, James, and
  Nurdin}}]{gough_quantum_2011-1}
\bibinfo{author}{\bibfnamefont{J.~E.} \bibnamefont{Gough}},
  \bibinfo{author}{\bibfnamefont{M.~R.} \bibnamefont{James}}, \bibnamefont{and}
  \bibinfo{author}{\bibfnamefont{H.~I.} \bibnamefont{Nurdin}}, in
  \emph{\bibinfo{booktitle}{2011 50th {{IEEE Conference}} on {{Decision}} and
  {{Control}} and {{European Control Conference}}}} (\bibinfo{year}{2011}), pp.
  \bibinfo{pages}{5570--5576}.

\bibitem[{\citenamefont{Gardiner and Zoller}(2004)}]{gardiner_quantum_2004}
\bibinfo{author}{\bibfnamefont{C.~W.} \bibnamefont{Gardiner}} \bibnamefont{and}
  \bibinfo{author}{\bibfnamefont{P.}~\bibnamefont{Zoller}},
  \emph{\bibinfo{title}{Quantum Noise}} (\bibinfo{publisher}{{Springer}},
  \bibinfo{year}{2004}).

\bibitem[{\citenamefont{Dum et~al.}(1992)\citenamefont{Dum, Parkins, Zoller,
  and Gardiner}}]{dum_monte_1992}
\bibinfo{author}{\bibfnamefont{R.}~\bibnamefont{Dum}},
  \bibinfo{author}{\bibfnamefont{A.~S.} \bibnamefont{Parkins}},
  \bibinfo{author}{\bibfnamefont{P.}~\bibnamefont{Zoller}}, \bibnamefont{and}
  \bibinfo{author}{\bibfnamefont{C.~W.} \bibnamefont{Gardiner}},
  \bibinfo{journal}{Phys. Rev. A} \textbf{\bibinfo{volume}{46}},
  \bibinfo{pages}{4382} (\bibinfo{year}{1992}).

\bibitem[{\citenamefont{Carmichael}(1993)}]{carmichael_open_1993}
\bibinfo{author}{\bibfnamefont{H.}~\bibnamefont{Carmichael}},
  \emph{\bibinfo{title}{An {{Open Systems Approach}} to {{Quantum Optics}}:
  {{Lectures Presented}} at the {{Universit\'e Libre}} de {{Bruxelles}},
  {{October}} 28 to {{November}} 4, 1991}}, Lecture {{Notes}} in {{Physics
  Monographs}} (\bibinfo{publisher}{{Springer-Verlag}},
  \bibinfo{address}{{Berlin Heidelberg}}, \bibinfo{year}{1993}).

\bibitem[{\citenamefont{Breuer and Petruccione}(2007)}]{breuer_theory_2007}
\bibinfo{author}{\bibfnamefont{H.-P.} \bibnamefont{Breuer}} \bibnamefont{and}
  \bibinfo{author}{\bibfnamefont{F.}~\bibnamefont{Petruccione}},
  \emph{\bibinfo{title}{The {{Theory}} of {{Open Quantum Systems}}}}
  (\bibinfo{publisher}{{Oxford University Press}}, \bibinfo{address}{{Oxford}},
  \bibinfo{year}{2007}).

\bibitem[{\citenamefont{Novoderezhkin et~al.}(2011)\citenamefont{Novoderezhkin,
  Marin, and van Grondelle}}]{novoderezhkin_intra-_2011}
\bibinfo{author}{\bibfnamefont{V.}~\bibnamefont{Novoderezhkin}},
  \bibinfo{author}{\bibfnamefont{A.}~\bibnamefont{Marin}}, \bibnamefont{and}
  \bibinfo{author}{\bibfnamefont{R.}~\bibnamefont{van Grondelle}},
  \bibinfo{journal}{Phys. Chem. Chem. Phys.} \textbf{\bibinfo{volume}{13}},
  \bibinfo{pages}{17093} (\bibinfo{year}{2011}).

\bibitem[{\citenamefont{Novoderezhkin et~al.}(2005)\citenamefont{Novoderezhkin,
  Andrizhiyevskaya, Dekker, and van Grondelle}}]{novoderezhkin_pathways_2005}
\bibinfo{author}{\bibfnamefont{V.~I.} \bibnamefont{Novoderezhkin}},
  \bibinfo{author}{\bibfnamefont{E.~G.} \bibnamefont{Andrizhiyevskaya}},
  \bibinfo{author}{\bibfnamefont{J.~P.} \bibnamefont{Dekker}},
  \bibnamefont{and} \bibinfo{author}{\bibfnamefont{R.}~\bibnamefont{van
  Grondelle}}, \bibinfo{journal}{Biophysical Journal}
  \textbf{\bibinfo{volume}{89}}, \bibinfo{pages}{1464} (\bibinfo{year}{2005}).

\bibitem[{\citenamefont{Raszewski and Renger}(2008)}]{raszewski_light_2008}
\bibinfo{author}{\bibfnamefont{G.}~\bibnamefont{Raszewski}} \bibnamefont{and}
  \bibinfo{author}{\bibfnamefont{T.}~\bibnamefont{Renger}},
  \bibinfo{journal}{J. Am. Chem. Soc.} \textbf{\bibinfo{volume}{130}},
  \bibinfo{pages}{4431} (\bibinfo{year}{2008}).

\bibitem[{\citenamefont{Roden et~al.}(2016)\citenamefont{Roden, Bennett, and
  Whaley}}]{roden_long-range_2016}
\bibinfo{author}{\bibfnamefont{J.~J.~J.} \bibnamefont{Roden}},
  \bibinfo{author}{\bibfnamefont{D.~I.~G.} \bibnamefont{Bennett}},
  \bibnamefont{and} \bibinfo{author}{\bibfnamefont{K.~B.}
  \bibnamefont{Whaley}}, \bibinfo{journal}{J. Chem. Phys.}
  \textbf{\bibinfo{volume}{144}}, \bibinfo{pages}{245101}
  (\bibinfo{year}{2016}).

\bibitem[{\citenamefont{Arsenault et~al.}(2020)\citenamefont{Arsenault, Yoneda,
  Iwai, Niyogi, and Fleming}}]{arsenault_vibronic_2020}
\bibinfo{author}{\bibfnamefont{E.~A.} \bibnamefont{Arsenault}},
  \bibinfo{author}{\bibfnamefont{Y.}~\bibnamefont{Yoneda}},
  \bibinfo{author}{\bibfnamefont{M.}~\bibnamefont{Iwai}},
  \bibinfo{author}{\bibfnamefont{K.~K.} \bibnamefont{Niyogi}},
  \bibnamefont{and} \bibinfo{author}{\bibfnamefont{G.~R.}
  \bibnamefont{Fleming}}, \bibinfo{journal}{Nat Commun}
  \textbf{\bibinfo{volume}{11}}, \bibinfo{pages}{1460} (\bibinfo{year}{2020}).

\bibitem[{\citenamefont{Arsenault et~al.}(2021)\citenamefont{Arsenault, Schile,
  Limmer, and Fleming}}]{arsenault_vibronic_2021}
\bibinfo{author}{\bibfnamefont{E.~A.} \bibnamefont{Arsenault}},
  \bibinfo{author}{\bibfnamefont{A.~J.} \bibnamefont{Schile}},
  \bibinfo{author}{\bibfnamefont{D.~T.} \bibnamefont{Limmer}},
  \bibnamefont{and} \bibinfo{author}{\bibfnamefont{G.~R.}
  \bibnamefont{Fleming}}, \bibinfo{journal}{J. Chem. Phys.}
  \textbf{\bibinfo{volume}{155}}, \bibinfo{pages}{054201}
  (\bibinfo{year}{2021}).

\bibitem[{\citenamefont{Tscherbul and Brumer}(2015)}]{tscherbul_partial_2015}
\bibinfo{author}{\bibfnamefont{T.~V.} \bibnamefont{Tscherbul}}
  \bibnamefont{and} \bibinfo{author}{\bibfnamefont{P.}~\bibnamefont{Brumer}},
  \bibinfo{journal}{J. Chem. Phys.} \textbf{\bibinfo{volume}{142}},
  \bibinfo{pages}{104107} (\bibinfo{year}{2015}).

\bibitem[{\citenamefont{Jeske et~al.}(2015)\citenamefont{Jeske, Ing, Plenio,
  Huelga, and Cole}}]{jeske_bloch-redfield_2015}
\bibinfo{author}{\bibfnamefont{J.}~\bibnamefont{Jeske}},
  \bibinfo{author}{\bibfnamefont{D.~J.} \bibnamefont{Ing}},
  \bibinfo{author}{\bibfnamefont{M.~B.} \bibnamefont{Plenio}},
  \bibinfo{author}{\bibfnamefont{S.~F.} \bibnamefont{Huelga}},
  \bibnamefont{and} \bibinfo{author}{\bibfnamefont{J.~H.} \bibnamefont{Cole}},
  \bibinfo{journal}{J. Chem. Phys.} \textbf{\bibinfo{volume}{142}},
  \bibinfo{pages}{064104} (\bibinfo{year}{2015}).

\bibitem[{\citenamefont{Baragiola et~al.}(2014)\citenamefont{Baragiola, Norris,
  Monta{\~n}o, Mickelson, Jessen, and
  Deutsch}}]{baragiola_three-dimensional_2014}
\bibinfo{author}{\bibfnamefont{B.~Q.} \bibnamefont{Baragiola}},
  \bibinfo{author}{\bibfnamefont{L.~M.} \bibnamefont{Norris}},
  \bibinfo{author}{\bibfnamefont{E.}~\bibnamefont{Monta{\~n}o}},
  \bibinfo{author}{\bibfnamefont{P.~G.} \bibnamefont{Mickelson}},
  \bibinfo{author}{\bibfnamefont{P.~S.} \bibnamefont{Jessen}},
  \bibnamefont{and} \bibinfo{author}{\bibfnamefont{I.~H.}
  \bibnamefont{Deutsch}}, \bibinfo{journal}{Phys. Rev. A}
  \textbf{\bibinfo{volume}{89}}, \bibinfo{pages}{033850}
  (\bibinfo{year}{2014}).

\end{thebibliography}

\end{document}